\documentclass[]{aastex}
\usepackage[]{emulateapj5}
\usepackage[]{epsfig}
\usepackage{graphicx}

\def\newline{\hfil\break}
\begin{document}
\title{Diffuse optical intracluster light as a measure of stellar tidal stripping: the cluster CL0024+17 at $z\sim $0.4 observed at LBT\footnotemark}
\footnotetext{Observations have been carried out using the Large Binocular Telescope at Mt. Graham, AZ. The LBT is an international collaboration among institutions in the United States, Italy, and Germany. LBT Corporation partners are The University of Arizona on behalf of the Arizona university system; Istituto Nazionale di Astrofisica, Italy; LBT Beteiligungsgesellschaft, Germany, representing the Max-Planck Society, the Astrophysical Institute Potsdam, and Heidelberg University; The Ohio State University; and The Research Corporation, on behalf of The University of Notre Dame, University of Minnesota, and University of Virginia.}
\author{E. Giallongo, N. Menci, A. Grazian, S. Gallozzi, M. Castellano, F. Fiore, A. Fontana, L. Pentericci, K. Boutsia, D. Paris, R. Speziali, and V. Testa}
\affil{INAF - Osservatorio Astronomico di Roma, via di Frascati
33, I-00040 Monteporzio, Italy}

\smallskip

\begin{abstract}
We have evaluated the diffuse intracluster light (ICL) in the central core of the galaxy cluster CL0024+17 at $z\sim 0.4$ observed with the prime focus camera (Large Binocular Camera) at the Large Binocular Telescope. The measure required an accurate removal of the galaxies light within $\sim 200$ kpc from the center.
The residual background intensity has then been integrated in circular apertures to derive the average ICL intensity profile. The latter shows an approximate exponential decline as expected from theoretical cold dark matter models where the ICL is due to the integrated contribution of light from stars which are tidally stripped from the halo of their host galaxies due to encounters with other galaxies in the cluster cold dark matter (CDM) potential.
The radial profile of the ICL over the galaxies intensity ratio (ICL fraction) is increasing with decreasing radius but near the cluster center it starts to bend and then decreases where the overlap of the halos of the brightest cluster galaxies becomes dominant. Theoretical expectations in a simplified CDM scenario show that the ICL fraction profile can be estimated from the stripped over galaxy stellar mass ratio in the cluster. It is possible to show that the latter quantity is almost independent of the properties of the individual host galaxies but mainly depends on the average cluster properties. The predicted ICL fraction profile is thus very sensitive to the assumed CDM profile, total mass and concentration parameter of the cluster.
Adopting values very similar to those derived from the most recent lensing analysis in CL0024+17 we find a good agreement with the observed ICL fraction profile.
The galaxy counts in the cluster core have then been compared with that derived from composite cluster samples in larger volumes, up to the clusters virial radius. The galaxy counts in the CL0024+17 core appear flatter and the amount of bending with respect to the average cluster galaxy counts imply a loss of total emissivity in broad agreement with the measured ICL fraction.
The present analysis shows that the measure of the ICL fraction in clusters can quantitatively account for the stellar stripping activity in their cores and can be used to probe their CDM distribution and evolutionary status.
\end{abstract}


\section{Introduction}

Galaxy clusters are high density regions in the Universe where we can test our knowledge of the physical processes governing galaxy formation and evolution. Massive galaxy clusters in particular contain a diffuse luminous component consisting of stars which are out of  galaxy halos in the intracluster environment. This intracluster light or ICL  was first detected by \citet{Zwicky51} and today we know that the ICL is an important component of the total stellar luminosity (e.g. \citet{Arnaboldi03,Arnaboldi04,Mihos05,Zibetti05,Krick07,Toledo11,Burke12,Guennou12}).
The first identification of the ICL was connected with the measure of very extended halos in the intensity profiles of the brightest cluster galaxies (BCGs), in excess with respect to the de Vaucouleurs ($r^{1/4}$) intensity profile for elliptical galaxies (\citet{Matthews64,Shombert88}). Estimates of the fraction of cluster light contained in the ICL in nearby clusters range from few \% up to 50\% depending on the cluster mass and redshift, where the highest value has been estimated in the Coma Cluster \citep{Bernstein95}.

Several models have been suggested to explain the origin of intracluster stars (see e.g. \citet{Tutukov11}).  Although various processes might contribute to some of the observed ICL it is widely accepted that stars in the external regions of galaxy halos are pushed into the intracluster environment by tidal stripping due to galaxy interactions (e.g. \citet{Weil97,Puchwein10,Rudick11,Martel12,Cui14} and references therein).

Recent N-body results \citep{Rudick11} predict that the ICL should contain a significant fraction ($\sim 10-40$\%) of the total stellar mass in clusters, in broad agreement with observations. Thus, the study of the ICL distribution in clusters can provide important information on the dynamical properties of the clusters and on their dark matter (DM) density profiles.

Different techniques have been adopted to measure the ICL in galaxy clusters since its definition is not unambiguous among observers. Moreover, different ICL classifications adopt different assumptions about the distribution of luminosity in galaxy halos. This results in different estimates of the intensity distribution of the ICL in the cluster cores,  as stated by \citet{Rudick11}. Indeed, some authors have measured the ICL below a given surface brightness threshold (e.g., \citet{Feldmeier04,Zibetti05}). Other authors have measured the ICL as an excess found in their model fits to galaxy halos
(e.g., \citet{Gonzalez05,Seigar07}). Because the study of ICL requires very deep, time-consuming observations, it is difficult to analyze the ICL distribution in different galaxy clusters within a single observational program (e.g., \citet{Mihos05,Krick06}).

To this end we began a long-term imaging program devoted to the analysis of the ICL in clusters of different physical properties at intermediate redshift. Wide field images are being obtained with the Large Binocular Camera (LBC, \citet{Giallongo08}), a prime focus camera at the Large Binocular Telescope (LBT). LBC is an ideal imager for this kind of study. Its high sensitivity due to the prime focus position at an 8m class telescope like LBT and its relatively wide field of view which allows a controlled background subtraction using regions far away from the cluster, provide an unique opportunity for the ICL detection.

We began the program with the well known massive cluster CL0024+17 at $z\simeq 0.4$ where 
an overwhelming amount of data and theoretical work are available in the literature (e.g. \citet{Tyson98,Broadhurst00,Czoske01,Treu03,Moran05,Jee07,Umetsu10}). A first detection of ICL for this cluster has been shown by \citet{Tyson98} who determined the ICL fraction with respect to the total light to be $\sim 15$\% within the 100 kpc region.  \citet{Jee10} has recently studied the ICL profile in this cluster, masking the core region and focusing the analysis on the external region $\sim 0.5$ Mpc looking for ICL signature of an external DM ring he claimed to be present from weak lensing analysis.

In the present paper we follow a different approach trying to remove the galaxy halo contribution with detailed profile fitting to derive an accurate ICL profile in the core region down to the BCG position and comparing this with simple theoretical predictions. 

In the following analysis all magnitudes are in the AB systems and all physical parameters have been computed adopting the standard $\Lambda$CDM model with $\Omega_{\Lambda}=0.7$, $\Omega_{0}=0.3$, baryonic density
$\Omega_b=0.04$ and Hubble constant $h=0.7$ in units of 100 km/s/Mpc.

\section{Data acquisition and reduction}

The data set was obtained with LBC at the LBT on Mount Graham in Arizona. LBC is a binocular camera installed at the prime focus of each of the two 8.4m telescopes. Each camera has an unvignetted field of view of 23$\times$ 23 arcmin$^2$ with a sampling of 0.226 arcsec/pixel.  LBC-Blue is optimized in the UV-B band and LBC-Red is optimized for the VRIZ bands.

The R-band image used for the present analysis has been obtained from frames  in the LBC-Red Sloan filter taken in 2011 November and 2012 January in photometric or clear sky conditions. The typical exposure time per frame was 120 s (which provided  4000 adu per pixel with a gain of 2$e^- adu^{-1}$ and a saturation for stars brighter than R=18.1 mag) and the final image is the coadding of several frames for a total of 2.73 hr of exposure time in an area which includes the core of the CL0024+17 cluster in a single chip, avoiding the analysis of inhomogeneities among different chips in the LBC field of view. The image has been calibrated using a zeropoint derived from Sloan r magnitudes of a set of Sloan Digital Sky Survey stars present in the field. We have verified that the LBC-Red Sloan R filter is very similar to the Sloan one and that any dependence on the (r-i)$_{Sloan}$ color 
is negligible. The rms uncertainty in the zeropoint is $\sim 0.07$ mag.

\subsection{Flat Fielding}
 The first step consisted of producing a coadded image of individual frames. The data were initially reduced using the LBC pipeline for imaging data: bias-subtraction, sky flat-fielding, and astrometric correction \citep{Giallongo08}.  

Specifically, we have applied skyflats (twilight flats) from blank field images taken the same night or the night just before, for the two runs of November $27-28$ and January 20 where the cluster data have been acquired. We have also applied superflats derived from different science images of deep, relatively empty fields, after removing all the detected objects down to the faintest limits allowed by the noise level. As emphasized in  the analysis of Gonzalez et al. (2005), changes in flat fields can affect the measure of any diffuse light along the chips. For this reason we have quantified the stability of the flat fields measuring the difference among various superflats and skyflats. Superflats changed from 2011 November to 2012 March  by up to 0.3\% but the flat accuracy remained confined at 0.06\% level since the adopted superflats have been computed using data within one to two days from the target observations. This accuracy would allow us to probe surface brightnesses 8.0 mag deeper than the sky level (R=21.1 mag arcsec$^{-2}$), corresponding to R=29.1 mag arcsec$^{-2}$. The same check has been applied to skyflats  measuring the difference between skyflats obtained in two consecutive nights. In this case, since we are interested in variations on scales of tens of arcsec which could affect the selection of ICL subregions, we have smoothed the frame with a Gaussian filter with a $\sigma$=2 arcsec to reduce the pixel to pixel noise level. The resulting difference is of the order of 0.05\% on a region of $1.7\times 1.7$ arcmin$^2$ around the cluster position.  In Figure 1 we have shown the stability of the superflat of the chip used for the cluster observations. The two superflats were obtained in two consecutive nights near the second cluster observing run. The superflat ratio in Figure 1(c) shows no specific spatial features. Time variation of the superflat on a day by day scale is thus confined on a spatial scale much smaller than the cluster angular size. The flat uncertainty is taken into account in the ICL estimate and shown as a horizontal threshold in the plot of the ICL profile shown in Section 3.

\subsection{Background subtraction}
The final coadded image has been obtained after equalization of each individual frame to the same value estimated in a region with poor contamination by faint sources. The resulting quality of the image corresponds to a resolution of FWHM$\simeq 0.73$ arcsec. The LBC pipeline also produces an rms map for each scientific image, directly from the raw science frame, as described in detail by \citet{Boutsia11}. These rms maps are used for the subsequent photometric analysis of the galaxies in the cluster.

The second step consisted of the objects detecting and masking. We ran Source Extractor (Sextractor,\citet{Bertin96}) to the coadded R-band image to create a source catalog. 

Concerning the scattered light from bright stars, this is an additive light which is often removed after careful fitting of the point-spread function (PSF) wings in bright saturated stellar images, as was done in, e.g.,  \citet{Gonzalez05} or \citet{Krick07}.
Since we are interested in measuring the ICL in the central core of the cluster ($2\times2$ arcmin$^2$), here we have followed  a different equivalent approach. 
After a first background estimate finalized to the selection of sources down to R=27 mag, we have masked all the sources including the overall $\sim 3\times2$ arcmin$^2$ cluster region shown in our Figure 2. 
We optimized parameters for source extraction of the faintest objects with a very low sky-$\sigma$ threshold. The galaxy detection was established at 2$\sigma$ level in an area within the FWHM.
This provided a first temporary catalog of galaxies detected down to $R\simeq 27$ and a Sextractor segmentation map used for subsequent object masking of the image.

Then we have estimated the background map using the Sextractor package. 
The choice of the mesh size (BACKSIZE) is of course important. If it is too small, the background estimation is affected by the presence of objects. Most importantly, the flux of the external regions of extended objects can be included in the background map. If the mesh size is too large, it cannot reproduce the small scale variations of the background. For this reason we have first used a backsize=256 pixel (LBC scale is 0.226"/pixel) for object selection and masking, then
we have used the much smaller value of backsize=32 pixel to absorb the extended wings of the bright saturated stars. Since the two nearest brightest stars are at
about 2.4 (the brightest) and 1.4 arcmin from the south edge of the cluster core shown in Figure 2, the backsize value adopted guarantees that in the midway, where we assume there is no detectable ICL, the background in empty regions is 0 within the rms noise level. In Figure 3 we show the enlarged background-subtracted image around the cluster core with the two saturated bright stars. We have tested that the results do not depend on the backsize value unless we use values $\gtrsim 256$ pixels. In the latter case a coarse sampling of the background map (e.g. $\lesssim 8\times 16$ mesh)  introduce stellar halo contamination in the background estimate. 
We show in Figure 3 an empty encircled region where the pixel intensity histogram has been computed. Its intensity distribution shown in Figure 4 suggests that we have removed any significant  contribution from bright stellar halos in the cluster core.
The background in the masked cluster region is finally interpolated from the neighborhood region.
The background-subtracted image has a 1$\sigma$ sky surface brightness limit of $R_{rms}\simeq 29.3$ mag arcsec$^{-2}$ estimated as in Boutsia et al. (2011).

We conservatively take into account any systematic error in the next ICL estimate due to the adopted background subtraction procedure producing different background-subtracted images. We adopt background maps with different backsize values (from 24 to 256 pixels),
ensuring that any residual background surface brightness in the selected regions was smaller than its rms value. These scatters in the background maps were taken into account when estimating error bars associated with the evaluation of the ICL profile described in the next section.

The resulting background-subtracted area around the CL0024+17 cluster core is shown in Figure 2.  The presence of a diffuse (green in the online journal) emission is clearly visible in the image. The diffuse emission surrounds the brightest galaxies extending outside their halos and including fainter and smaller dwarf galaxies. Sharp edges are clearly visible in the external region of the cluster core outside about 50 arcsec from the BCG. 

\section{ICL estimate}

To evaluate the measure of diffuse emission in the cluster core, we need to remove the galaxy halo contamination. Previous evaluations on the same cluster by \citet{Jee10} were mainly focused on external regions up to $R\sim 0.5$ Mpc to probe the possible presence of an external DM ring  which has then been questioned by subsequent lensing analysis \citep{Umetsu10}. For this reason, almost all of the central cluster core was masked in the previous analysis. Here we follow an opposite approach, trying to remove the galaxy halo contribution by detailed profile fitting of the galaxies present in the cluster core, both cluster members or foreground and background/lensed.

We applied the Galfit package \citep{Peng10} for profile fitting all the galaxies found by Sextractor down to $R\simeq 27$ in the cluster core ($R\lesssim 200$ kpc from the BCG). Galfit is one of the most accurate softwares, allowing us to fit the galaxy profiles of ellipticals, spirals, as well as irregular galaxies adopting parametric functions like the S{\'e}rsic, Moffat, King, Ferrer, etc., profiles.

We used S{\'e}rsic power-law  models which can mimic radial distributions of different galaxy types ranging from spirals to ellipticals. A PSF estimated using relatively bright stars present near the cluster was convolved with the intrinsic profiles. When the power-law index $n$ is large, it has a steep inner profile and an extended outer wing. On the contrary, when $n$ is small, it has a shallow inner profile and a sharp drop at large radius. The traditional de Vaucouleurs and exponential disk profiles are specific cases with  $n=4$ or 1, respectively. It is clear from Figure 4 of \citet{Peng10} that profile fitting with  large ($n>5$) power-law indices imply very extended halos which in our case can emulate the presence of any extended background around bright galaxies, for this reason best fits were considered acceptable for $n<4.5$. 

The objects were fitted in groups as a tradeoff among different requirements.
A relatively large area including more galaxies was required for a better estimation of the local diffuse background. The group should not be too large to allow the best sensitivity in the fit of the more numerous faint objects compared to the brighter ones. Finally, the computer time needed for convergence is a further limiting parameter.

 We have first fitted the central region of the cluster core: the original, fitted and residual images are shown in Figure 5(a), (b), and (c). The region for the simultaneous galaxy fit was selected to sample the connected diffuse light on the basis of the isophotal connection limited to  27.4 mag arcsec$^{-2}$.

Previous analyses by Gonzalez et al. 2005 tried to fit the external BCG halos of lower redshift, well relaxed clusters with a two-component de Vaucouleurs profile (instead of a single S{\'e}rsic profile, with an index often greater than 4) and identify the external component as being due to the ICL contribution. To fit the external light we adopted a "modified Ferrer profile"  \citep{Peng10} which consists of a central power-law shape followed by an external cutoff whose truncation sharpness can be tuned by a free parameter.
In addition to the standard radial profiles, azimuthal shape functions like Fourier and bending modes were used to add azimuthal perturbations to the radial profiles. These azimuthal functions were adopted specifically to account for lensed, distorted galaxy images or tidal tails, as well as to model the irregular shape of the observed diffuse light contours at various intensity levels.

All the galaxies have been fitted with S{\'e}rsic profiles with resulting best fit values $n<4$ with the exception of the bright A, B, and C galaxies indicated in Figure 5(a). For these galaxies a single component analysis gave index values up to $n\sim 6$ and a significant ring appeared in the halo residuals together with a negative intensity hole in the center. Although the fitting of the region shown in Figure 5(a) was still acceptable with a reduced $\chi^2=1.7$, following Gonzalez et al. (2005), we included a second component in the fit which, with few more degrees of freedom, provided a significant reduction of the $\chi^2$ by about $-9500$. The resulting fit gave $n<4.5$ for all the double components of the three mentioned galaxies and a small and uniform residual where holes and rings disappeared (Figure 5(c)). The resulting best fit gave a reduced $\chi^2=1.5$ and the residual map shown in Figure 5(c) gave an average value$=-0.003$ adu and a noise rms$=0.067$. The best fit S{\'e}rsic values for the A, B, and C galaxies are shown in Table 1 as an example. 
The best fitting modified Ferrer profile values are: central brightness 24.3, $\alpha=9.2$ (outer truncation sharpness), $\beta=1.4$ (central slope), an axis ratio of 0.8, an outer truncation radius of 231 arcsec, a position angle of 67 deg, bending coefficients 298.8,4406, and $-8269$, Fourier terms F1, F3, F4 0.3, $-0.1$, and 0.08, respectively. The effects due to the bending modes and Fourier terms can be seen in Figure 5(b) where the low-intensity profile (green) is curved toward the lower left corner with two main blobs on the left side. 
More external regions have been fitted in a similar way. This ensured that any residual halo intensity of each galaxy has been excluded from the ICL analysis.

At the center of the cluster where the four brightest galaxy halos overlap, the ICL measure mainly depends on the interpolation of the background shape estimated in the more external region. We emphasize that at the low resolutions of ground-based observations it is difficult to recover the true, intrinsic model parameters of the intensity profile in the galaxy cores. However, what is important for the estimate of the ICL is the best fit of the galaxy profile which minimizes the residuals. 

The overall best-fit solution is shown in Figure 6(a) where the intensity distributions of all the fitted galaxies are shown. In the same Figure 6(a)  it is possible to evaluate the contribution to the apparent diffuse light by the overlaps of the galaxy halos which is relevant around the cluster brightest galaxies. Figure 6(b) shows the image in Figure 6(a) subtracted from the original image, where the smoothed ICL is shown as a residual to the galaxy profile fitting. On the basis of this procedure the ICL shown in Figure 6(b) could be considered as a lower limit if some physical radial truncations were present in the galaxy intensity profiles. The ICL shows a main structure elongated in the direction northwest-southeast (NW-SE) and a clear substructure in the southwest (SW) side of the image. 

The ICL distribution extends at least up to 200 kpc (the scale is $\sim 5.3$ kpc/arcsec)  from the barycenter located at R.A.=00 26 35.478 decl.=17 09 43.6.  The center is few arcsec from the X-ray (R.A.=00 26 36.3 decl=17 09 46) and DM (R.A.=00 26 33.37 decl.=17 09 41.68, \citet{Umetsu10} centers.

Various aperture magnitudes with their associated errors were computed using the Sextractor package. The resulting circular intensity profile is shown in Figure 7. It is interesting to note the resulting nearly exponential profile of the ICL which extends up to 140 kpc. Fitting a linear relation in the $R=30-140$ kpc region gives the best fit result shown in Figure 7: ICL$=a+R/R_0$ with $a=24.80\pm 0.03$ and $R_0=48\pm 1$ kpc.  At 150 kpc a small bump is present that is associated with the SW substructure and the SE elongation and
at larger radii the profile appears to resume an exponential behavior. It is not obvious to find out such an exponential behavior from a region that is influenced by de Vaucouleurs profiles of early-type galaxy halos. However in a CDM framework where the ICL production is mainly due to stellar tidal stripping from galaxy halos, an exponential behavior is expected for the stellar mass lost by galaxy halos which is roughly proportional to the ICL intensity in the red bands, as shown in the next section. Error bars in Figure 7 take into account possible systematic errors in the background subtraction as analyzed in Section 2. The horizontal line shows the level where uncertainties in the flat fielding procedure affect the ICL measure.

It is possible at this point to evaluate the intensity ratio profile of the ICL, ${\cal F}_{ICL}$, defined as the ratio in circular apertures between ICL and galaxy intensities. The ratio is shown in Figure 8 in differential form after removal of galaxies whose available spectroscopic or photometric redshifts (see e.g. \citet{Czoske01,Treu03,Smith05}) are not compatible with the cluster membership. Uncertainties in the evaluation of the cluster membership are  small  for $R_{AB}<21.5$ since most of the galaxies in the cluster core (Figure 6(a)) have spectroscopic redshifts (33/43) and the remaining accurate photometric $z$. For increasing magnitudes,
the use of photometric redshifts with typical errors of 0.1 \citep{Smith05} gradually increases the uncertainties in the cluster member assignment. For $R_{AB}>23$ we assume all the galaxies in the cluster core as cluster members. Since they are faint and small, any uncertainty in their membership fraction should change the estimate of the ICL fraction by $\sim 10$\%.

The ratio profile in annular regions appears to increase as the radius decreases from 200 kpc down to $70-80$ kpc reaching a peak value of ${\cal F}_{ICL}\sim 40$\%. At lower radii a bending followed by a decrease is present. The decrease is due to the presence of three among the BCGs  whose extended halos overlap in the small central region. Thus, any evidence of diffuse light in the cluster center relies on the extrapolation at $R<50$ kpc of the modeled background gradient behind the central galaxy profiles. The integrated fraction within $R\sim 100-150$ kpc
 is  ${\cal F}_{ICL}\simeq 23$\%. This integrated value is in between the first \citet{Tyson98} 15\% estimate and the 35\% value derived by \citet{Jee10}. Our integrated value is that expected on average by numerical simulations of cluster evolution in the CDM scenario from clusters of similar mass (e.g. \citet{Henriques10,Rudick11,Martel12}). The differential ICL profile, however, can give more detailed information on the shape of the cluster potential well, as outlined in the next section.

\section{ICL predictions from tidal stripping}

Here we show how basic quantities related to the shape of the cluster potential wells can be extracted from the 
observed radial dependence of the ratio ${\cal F}_{ICL}$ in Figure 8.  In fact, a straightforward, though simplified, analytical 
computation of the ICL fraction ${\cal F}_{ICL}(x)$ can be worked out starting from the canonical expression 
(see \citet{King62,Taylor01}) for the tidal radius of a satellite galaxy (with initial mass $m_s$, 
halocentric radius $r$, and angular velocity $\omega$) in a potential $\phi(r)$: 
\begin{equation}
r_t=\Bigg({G\,m_s\over \omega^2-{d^2\phi\over dr^2}}\Bigg)^{1/3}
\end{equation}

The above expression identifies the tidal radius with the distance to the saddle point in the potential interior to the satellites orbit, since this is the point at which the radial forces on a test particle cancel out (\citet{vonh57,King62,Binney87}). Here we shall focus on the effect of the external potential on the stellar distribution of the orbiting galaxies; note in our approach we deliberately neglect other processes which may be relevant for stripping, such as the cumulative effects of gravitational two-body interactions (\citet{Gallagher75,Richstone75,Merritt83}; for an N-body investigation including galaxy harassment see \citet{Moore96}), or pre-processing (\citet{Mihos04}; see also \citet{Fujita04}). Our approach is to estimate whether the effect of the tidal field of the cluster gravitational potential is sufficient to explain the observed ICL profile. An investigation of such effect on galaxy orbits and stripping has been performed by \citet{Taylor01}; these authors, however, focussed on the detailed description of the possible orbital evolution of galaxies, but do not compute the effects on the statistical distribution of galaxies and ICL for different profiles. Previous computations investigating the tidal effects of the cluster potential (starting from \citet{Miller83}) have been performed through N-body simulations;  in particular \citet{Bekki03}) simulations were aimed at determining the effect of such process on the dwarf population of the Fornax cluster, finding that such tidal effects may appreciably affect the stripping and the final properties of such dwarf galaxies. Here we analytically compute - under simplified assumptions - the effect of the cluster tidal field on the stripping of stars for a generic gravitational potential, to investigate whether such process can account for our observational results.

Let us assume that satellite galaxies move on circular orbits (we shall come back on this point later). We write the dark matter density profile 
of  the cluster in the form $\rho(r)=\rho_0\,f(x)$, where $\rho_0$ is a central density and $x\equiv r/r_c$ is the distance from 
the cluster center in units of a scale radius $r_c$; 
for the standard \citet{Navarro97} profile, this is defined in terms of the viral radius $r_v$ and of the concentration parameter $c$ as 
$r_c=r_v/c$. An analogous notation for the satellite yields for the satellite mass $m_s=4\,\pi\,r_{cs}^3\,\rho_{0s}\int^{x_s}_0\,f(x'_s)\,x_s'^2\,dx'_s$
where $x_s\equiv r_s/r_{cs}$, and $r_s$ and $r_{cs}$ are  the effective boundary radius and the scale radius of the satellite galaxy, respectively.
Then Equation (1) becomes 

\begin{equation}
r_t={\sigma_{s}\over \sigma}\,x\,r_c\,A(x)
\end{equation}
where we have defined an effective velocity dispersion in the cluster $\sigma\propto\rho_0\,r_c^2$ and similarly in the satellite galaxy (in analogy with the isothermal case) and

\begin{equation}
A(x)\equiv\Bigg\{{[I(x_s)/x_s]^{3/2}\over [I(x)/x]^{1/2}\,\bigg[ [2\,I(x)/x]+x\,{d\over dx} [I(x)/x] \bigg]}\Bigg\}^{1/3}~, 
\end{equation}
where we have defined the function $I(x)\equiv \int^{x}_0 f(x')\,x'^2\,dx'$.  Details are shown in the Appendix.

Note that in the isothermal case is $f(x)=x^{-2}$, so that $I(x)=x$ and Equation (2) reduces to the 
expression $r_t=(\sigma_{s}/\sigma)\,r/2^{1/3}$. In this respect, the function $I(x)/x$ in Equation (3) represents the deviation from the isothermal case.
 
To estimate the amount of stars lost by a satellite galaxy at a distance $r$ from the center (in a circular orbit), we compute the stellar mass
that lies beyond the tidal radius (5), assuming an exponential form for the initial stellar mass distribution: 
\begin{equation}
m_{lost}=\int_{r_t}^{\infty}2\,\pi\,\Sigma_0\,e^{-{\xi\over r_d}}\,\xi\,d\xi
\end{equation}
where $r_d$ is the disk exponential scale length and $\Sigma_0$ the central surface density.
We now substitute  expression (2) for $r_t$ in Equation (4) obtaining
 \newline
\begin{eqnarray}
 m_{lost} & = & m_*\,G(x,\lambda,c)  \nonumber   \\
 G(x,\lambda,c) & \equiv & \Bigg[1+{x\,A(x)\over \lambda\,c}\Bigg]\,exp{\Bigg[-{x\,A(x)\over \lambda\,c}}\Bigg]
\end{eqnarray}
where $m_*$ is the initial stellar mass of the satellite galaxy and $\lambda=r_d/r_{vs}$ is the spin parameter or equivalently the ratio between its scale length and its virial radius \citep{Mo98}.
For circular orbits, the contribution to the intracluster stellar mass from all galaxies at a distance in the range  $(x, x+dx)$ from the cluster center can
be obtained summing up the contribution of all galaxies orbiting at the radial distance  $x$ from the 
cluster center, i.e., 
\begin{eqnarray}
M_{lost}(x) & =& \int dm_*N(m_*)\,w(x)\,x^2\,m_*\,G(x) \\
& = & \overline{m_*}\,w(x)\,x^2G(x) \nonumber 
\end{eqnarray}
\newline
where $w(x)$ - for any given stellar mass - is the number density of galaxies  between $x$ and $x+dx$, and $\overline{m_*}\equiv \int\,dm_*\,N(m_*)\,m_*$ is the average stellar mass before stripping. The analogous expression for the total initial stellar mass in galaxies before stripping is 
\begin{equation}
M_{*}(x)=\int dm_*N(m_*)\,w(x)\,x^2\,m_*= \overline{m_*}\,w(x)\,x^2.
\end{equation}
In particular, in the simplifying assumption of constant mass to light ratio, 
the ratio ${\cal F}_{ICL}$ of intra-cluster to galactic light assumes a simple form 
\begin{equation}
{\cal F}_{ICL}(x)\approx {M_{lost}(x)\over M_*(x)-M_{lost}(x)}={G(x)\over 1-G(x)  }. 
\end{equation}

In the context of our simple physical description of tidal stripping, the latter expression shows that the ratio ${\cal F}_{ICL}$ is independent of the physical and statistical properties of the satellite population. Thus the interesting result is that the ratio  {\it mainly depends on the gravitational profile of the cluster}. Thus, the ICL fraction can be used as a further probe of the large scale cluster physical properties.

In fact, to compare with observations, Equation (8) should be projected on the direction perpendicular to the line-of-sight as shown in the Appendix (Equation (A.7)).  We remark that the above computation has been derived assuming circular orbits. However, the distribution of 
halo circularities in simulated clusters is peaked at values $\epsilon\approx 0.8$ (see, e.g., \citet{Ghigna98}), so that a more realistic computation can be performed assuming that - for any radial shell at a distance $r$ from the cluster center - galaxies oscillate within a distance $\Delta r$
from $r$. In such a case we can perform a numerical computation adopting, for the angular velocity $\omega$, the proper value at any point of the orbit, and
assuming different values for $\Delta r$ so as to explore the possible circularities between 0.7 and 1. The results from such a numerical computation
are almost identical to those obtained analytically from Equation (8), which we then use to interpret the observational radial distribution ${\cal F}_{ICL}$. 

When compared to our observational results for ${\cal F}_{ICL}$ (Section 3),  Equation (A.7) can be used to probe  the shape of the cluster 
potential wells. In particular, we can probe the effect of assuming different forms for the function $f(x)$ defining the shape of the 
density distribution $\rho(x)=\rho_0\,f(x)$. The procedure is as follows. We chose different forms for $f(x)$ and for the 
concentration parameter $c$ , which determines the function $G(x)$ entering the ICL ratio ${\cal F}_{ICL}(x)$ given in Equations 8 and A.7 through the expression in Equation (3). This is compared to the observed radial dependence of the ICL ratio determined in Section 3, to estimate which of the input forms for $f(x)$ provides the best match to the observed run. We focus on the cored isothermal form $f(x)=1/(1+x^2)$, and on the \citet{Navarro97} form $f(x)=1/x(1+x^2)$, and we always assume a fixed canonical value for the spin parameter $\lambda=0.1$.
We take for the virial radius of the cluster the observed value $r_v=1.6/h$ Mpc; for fixed virial radius, the radial coordinate $x=r/r_c=c\,r/r_v$ depends on the concentration parameter $c$ for which we explore different values for each of our assumed form of $f(x)$. The results of the comparison are shown in Figure 8 for 
both the isothermal  and  NFW profiles. The best agreement is obtained for a NFW profile with $c=9$; interestingly, this is very close to the value $c=9.2$ obtained from the combined strong-weak lensing analysis by \citet{Umetsu10}. Note that we do not attempt to fit the inner region of the ${\cal F}_{ICM}(r)$; the observed turning down of the profile 
is affected by the accretion effect of the central galaxy which is by no means included in our analytical computation.
Finally, in our computation we have also assumed the satellite galaxies to have the same density profile of the galaxy cluster, i.e. $c=c_s=9$, which is reasonable for this specific cluster. However, we have checked that - in the \citet{Navarro97} case - adopting a difference $\left|{c-c_s}\right|\sim 4\div 5$  yields only minor changes to the results obtained assuming the same density profile (in fact, we can recast its effect in terms of a normalization of $A(x)$ in Equation (3) differing by less than 5\% from our reference cases).

Nevertheless, the remarkable agreement found between the ICL fitting technique described above and the measurements from weak lensing shows that, despite the approximations adopted in our analytical computation, the radial dependence of the ICL distribution can constitute a complementary probe for the shape of the clusters potential wells.

\section{ICL predictions from galaxy counts}

If the spread of ICL in the cluster core is due to the diffusion of unbound stars stripped from the halos of satellite galaxies, the resulting stellar mass loss is expected to alter 
the luminosity of cluster galaxies. A sign of this effect should be found in the shape of the galaxy luminosity function (LF) in the cluster core.

For this reason we built binned number counts for all the fitted galaxies which have then been removed for the ICL measure. The number counts are shown in Figure 9 and appear 50\% complete down to $R_{AB}\simeq 26.5$. For this reason we limited the statistical analysis to $R_{AB}=26$. We derived absolute magnitudes adopting the appropriate redshift and $k-$corrections estimated from \citet{Fukugita95} for E-Sab galaxies.

We fitted a Schechter function to the observed counts which have the same shape of the galaxy luminosity function. This assumption implies that all the galaxies shown in the counts are at the same cluster redshift. This is justified by the fact that in order to compute the ICL fraction with respect to the total galaxy light we have already removed from the catalog the foreground/background galaxies selected on the basis of their colors and spectroscopic redshifts, where available.

However it is well known that the Schechter function generally underestimates the abundance of very bright galaxies ($M_R<-22.5$), which is mainly due to the presence of the BCGs (e.g \citet{Christlein03}). For this reason we fitted them with a second Schechter component.

For the scope of the present work, we applied a simple $\chi^2$ fitting to the data points shown in Figure 9. We adopted the sum of two Schechter functions. The best fit values found are $M^*\simeq -21.65$ and $\alpha\simeq 3.7$ for the BCGs  component and  $M_{br}^*=-22.2$, $\alpha \simeq -1.2$ for the main component. We emphasize that the atypical best fit values of the former component have been obtained, forcing a Schechter shape to the brightest data point.

We have compared the best fit parameters of the main counts with that derived from composite samples of clusters at various redshifts. More specifically we have selected the composite luminosity functions derived by \citet{Harsono09}. Their sample is composed by archival Hubble Space Telescope/Advanced Camera for Surveys images of six clusters in the redshift interval $z\simeq 0.14-0.4$, including also the present cluster. They derive a characteristic luminosity $M^*\simeq -22\pm 0.3$ in the R band and a faint end slope $\alpha\simeq -1.33\pm 0.03$ for galaxies within the virial radius, i.e. in a much larger volume than sampled in the core of CL0024+17.

\citet{Alshino10} derived stacked luminosity functions of XMM–LSS clusters of Class 1 at different average redshifts. In particular at redshifts $z\simeq 0.3$ they find $M^*\simeq -21.99\pm 0.25$ and $\alpha\simeq -1.47\pm 0.07$. Their slope is steeper, probably due to the larger area sampled for their clusters as noted by the same authors. We retain both slopes to enclose the uncertainties present so far at the faint end.

Although there is a good agreement on the value of the characteristic luminosity among CL0024+17  and the two cluster samples, the faint end slope in our cluster appears somewhat flatter. A speculative explanation for this flattening is given in terms of a stellar tidal stripping from the halos of galaxies in the cluster core. 

The two luminosity function shapes are shown in Figure 9 in terms of surface counts normalized to the CL0024+17 counts at $M^*=-22.2$. The bright Schechter component found in CL0024+17 has then been added to the two counts.

Since galaxies of intermediate mass/luminosity are those which lose a significant amonut of stellar mass from their halos due to tidal stripping, they dim progressively their luminosity as clusters relax. Thus it is interesting to compare the CL0024 galaxy emissivity $\epsilon_{CL0024}$ with that predicted by the two cluster luminosity functions. Specifically we have computed the quantity
\begin{equation}
(\epsilon_{clf}-\epsilon_{CL0024})/\epsilon_{CL0024}
\end{equation}
for the two composite luminosity functions, where the integration in luminosity has been done in the representative interval $M=-24\div -15$.

It is interesting to note that the galaxy emissivity in the CL0024 core (within 200 Kpc) is smaller than that predicted on the basis of the average cluster luminosity functions and the fractional loss of galaxy emissivity ranges from 19\% to 37\% for the two slopes, respectively. Thus, the fractional galaxy emissivity missing in the CL0024 core appears to be consistent with the fractional emission of the ICL, $\simeq 23$\% of the galaxy emission.

This result reinforces the scenario where ICL in cluster cores is produced by stars stripped from the halos of their parent galaxies in the cluster potential well, as described in the previous section. The stripping produces a flattening of the  faint end counts or luminosity functions in the cluster core with respect to the average shape derived from cluster galaxy counts within the virial radius. Thus the average cluster luminosity functions derived by  \citet{Alshino10} and \citet{Harsono09} appear steeper than our LF derived in the CL0024 core since they are derived in  much larger clusters volumes, up to the virial radius (e.g. of the order of megaparsec). In the latter case the LFs are steeper since they are dominated by galaxies located far from the central 200 kpc region where tidal stripping is more effective and ICL is more intense. In other words, we predict a faint LF slope which should flatten considering volume shells approaching the cluster center. The amount of flattening observed in the CL0024 luminosity function is quantitatively consistent with the measured ICL within a radius of 200 kpc.

\section{Summary}
We have derived an estimate of the diffuse ICL in the cluster CL0024+17 at $z\simeq 0.4$ to connect this quantity and its spatial distribution with the dynamical properties of the cluster core mainly due to the action of the stellar tidal stripping.

The main objective has been reached through an accurate evaluation of the systematics introduced in the flat fielding and background subtraction procedures, and after careful removal of the galaxies light in the cluster center. This has been obtained through profile fitting using the galfit software package. Only the sum of the galaxy profiles has then been subtracted from the original image and the residual intensity distribution obtained in this way was considered representative of the ICL spatial distribution.
The main results can be summarized as follows:
\begin{itemize}
\item
The ICL distribution due to the stripping action of the brightest galaxies in the cluster core extends at least up to 200 kpc from the center.
\item
The radial intensity profile of the ICL derived from circular apertures has an exponential behavior. A small deviation is present at $R=150$ kpc due to the action of a distinct group of galaxies. An exponential decrease is consistent with the expectation of theoretical CDM models where intracluster stars are produced by tidal stripping in galaxy halos.
\item
After removing foreground galaxies on the basis of the available spectroscopic or photometric redshifts, we have estimated an integrated ICL fraction of $\simeq 23$\% of the overall galaxy light
within 150 kpc.
\item
The average radial profile of the ICL fraction increases with decreasing radius reaching a maximum value $\sim 40$\% at $R\sim 70-80$ kpc.  At smaller radii a bending is present followed by a significant decrease due to the overlap of the halos of the cluster brightest galaxies.
\item
Simple predictions in CDM scenarios show that the ratio between the total stellar mass lost by stripping over the stellar mass present in galaxies is almost independent of the physical properties of the generic galaxy hosting stripping action. The ratio mainly depends on the global cluster properties such as e.g., total mass, and DM profile shape. Adopting the mass value of $M=1.7 \times 10^{15}$ M$_{\odot}$ and a Navarro-Frenk-White DM profile with concentration parameter $c_{vir}= 9$, very close to the values derived from extensive lensing analysis on this cluster, we predict an ICL fraction profile in broad agreement with the observational trend for $R>70$ kpc.
\item
We have computed the galaxy counts in the core of the cluster ($R<200$ kpc).  Fitting two Schechter laws, one for the brightest bin populated by the dominant brightest galaxies in the core and one for the remaining sample, we derived a characteristic absolute magnitude $M^*\simeq -22.2$ for the main component, consistent with that estimated from composite cluster samples at similar redshifts. The faint slope, however, appears definitely flatter, $\alpha \simeq -1.2$, with respect to the average slope derived from composite cluster samples ($\alpha \sim -1.3$ to $ -1.5$). 
\item
Since the composite cluster samples are derived on much larger volumes up to their virial radius, we have tested whether stellar tidal stripping can be responsible for the observed bending of the counts in our cluster core. 
Normalizing composite average counts (or equivalently luminosity functions)  to the value derived for CL0024+17 at $M^*\simeq -22.2$, we have computed the relative difference of total emissivity produced by the composite and CL0024+17 counts. This difference amounts to 19\%$-37$\% for $\alpha=-1.3, -1.5$, respectively. The emissivity "lost" in CL0024+17 appears in broad agreement with the ICL fraction we have measured in the core, $23$\%, suggesting that the stripping activity, bending the galaxy counts, can quantitatively explain the observed  ICL fraction. 
\end{itemize}

This pilot study will be applied to more clusters at different redshifts to explore the dependence of the ICL fraction on the cluster physical (e.g. mass and concentration) and evolutionary (e.g. relaxation status) properties. A large amount of statistics will enable us to follow on average the stellar stripping activity during the evolution of the cluster cores.

\acknowledgments
We thank the referee for detailed comments which helped us to improve the presentation of the paper and the robustness of our results. We also thank T. Treu for useful discussions.


\appendix
In this appendix we derive more explicitly some expressions used in the text.
Defining the function $I(x)$ as in the text, Equation (1) can be rewritten as
\begin{equation}
r_t=\Bigg({4\,\pi\,G\,\rho_{0s}\,r^3_{cs}\,r^2\,I(x_s)\over 2\,v_c^2(r)+r\,{dv_c^2\over dr}}\Bigg)^{1/3}~,
\end{equation}
where we have expressed the denominator in Equation (1)  
in terms of the square of the cluster circular velocity $v^2_c(r)=G\,M(<r)/r$. Writing the latter as 
$v^2_c=4\,\pi\,G\,\rho_0\,I(x)\,r_c^3/r$ we obtain
\begin{equation}
r_t=\Bigg({\rho_{0s}\,r^2_{cs}\over \rho_0\,r_c^2}\,r^2\,r_s {I(x_s)/x_s\over 2\,I(x)/x+x\,{d I(x)/x\over dx}}\Bigg)^{1/3}~.
\end{equation}
This can be simplified defining an effective velocity dispersion  as $\sigma\propto\rho_0\,r_c^2$ (in analogy with the isothermal case). 
The satellite size $r_s$ can be related to the density profile assuming that at the tidal radius is  $\overline{\rho(r_s)}=\overline{\rho(r)}$, which yields 
$r_s/r=\sigma_s/\sigma\,[ I(x_s)\,x/I(x)\,x_s]^{1/2}$; then Equation (2) reads
\begin{equation}
r_t={\sigma_{s}\over \sigma}\,x\,r_c\,A(x)
\end{equation}
where $A(x)$ is defined in the text (Equation (3)).

Assuming an exponential form for the initial stellar mass distribution, the mass lost can be written as
\begin{equation}
m_{lost}=\int_{r_t}^{\infty}2\,\pi\,\Sigma_0\,e^{-{\xi\over r_d}}\,\xi\,d\xi=m_*\Big(1+{r_t\over r_d}\Big)\,e^{-{r_t\over r_d}}
\end{equation}
where $r_d$ is the disk exponential scale length, and we have expressed the central disk surface density $\Sigma_0$ in terms of the initial stellar mass  of the galaxy satellite $m_*=\int_0^{\infty}2\,\pi\,\Sigma_0\,e^{-{\xi\over r_d}}\,\xi\,d\xi$.

We relate the disk size $r_d$ to the virial radius of the satellite galaxy $r_{vs}$ assuming angular momentum conservation during the formation of the disk. 
Assuming equal values for the baryon-to-DM angular momentum and mass ratios,  yields $r_d=\lambda\,r_{vs}$  (\citet{Mo98}), 
where $\lambda$ is the spin parameter of the DM halo, for which we take the standard average value $\lambda\approx 0.1$. Then Equation (4) yields 
\begin{equation}
m_{lost}=\int_{r_t}^{\infty}2\,\pi\,\Sigma_0\,e^{-{\xi\over r_d}}\,\xi\,d\xi=m_*\Big(1+{r_t\over \lambda\,r_{vs}}\Big)\,e^{-{r_t\over \lambda\,r_{vs}}}.
\end{equation}

We now substitute  expression (2) for $r_t$ in the latter equation, substitute $r_c=r_v/c$ for the cluster core radius, 
and use the standard scaling relations $\sigma\propto \big[GM/(M\overline {\rho})^{1/3}\big]^{1/2}$ 
and $r_v\propto \big[M/\overline {\rho}\big]^{1/3}$ for both the cluster and 
 the satellite galaxies. This leaves us with Equation (5) shown in the text
 \newline
\begin{eqnarray}
 m_{lost} & = & m_*\,G(x,\lambda,c)  \nonumber   \\
 G(x,\lambda,c) & \equiv & \Bigg[1+{x\,A(x)\over \lambda\,c}\Bigg]\,exp{\Bigg[-{x\,A(x)\over \lambda\,c}}\Bigg]
\end{eqnarray}

Finally,  to compare with observations, we shall use  Equation (8) in the text projected on the direction $x_{\bot}$ perpendicular to the line-of-sight. This follows from integrating 
Equation (8) in the text over the perpendicular direction $x_{\|}$ after weighting with the density of galaxies $w(x)\,x^2$ in any radial shell 
 $x=\sqrt{x_{\bot}^2+x_{\|}^2}$. 
\begin{equation}
{\cal F}_{ICL}(x_{\bot})={\int_0^{x_{\|max}}\,d  x_{\|} x^2\,(x_{\bot},x_{\|})\,w[x(x_{\bot},x_{\|})]\,G[x(x_{\bot},x_{\|}),\lambda,c ]\over 
\int_0^{x_{\|max}}\,d  x_{\|} x^2\,(x_{\bot},x_{\|})\,w[x(x_{\bot},x_{\|})]\,\Big[1-G[x(x_{\bot},x_{\|}),\lambda,c ]\Big]
 },
\end{equation}
where $x_{\|max}=\sqrt{c^2-x_{\bot}^2}$.

{}

\clearpage
\begin{center}{\bf Table 1}\\
\vskip 0.5truecm
\begin{tabular}{|l|l|l|l|}
\hline
Galaxy$^a$ & mag$_R$ ($\pm 0.05$) & R(pixels) ($\pm 0.1$)& S{\'e}rsic Index n ($\pm 0.06$)\\
\hline
 A$_1$    & 19.16  &   6.8   &   3.87     \\
 A$_2$    & 19.94  &  16.9   &  1.35     \\
 B$_1$    & 19.42  &   12.7   & 1.24     \\
 B$_2$    & 19.73  &   4.1   &   4.45     \\
 C$_1$    & 19.48  &   5.7   &   2.37     \\
 C$_2$    & 19.92  &  17.1   &  1.06     \\

\hline
\end{tabular}
\\
$^a$ Galaxies A, B, and C are labeled in Figure 5(a).
\end{center}

\clearpage

\begin{center}
\vspace{0.cm}
\scalebox{0.4}[0.4]{\rotatebox{0}{\includegraphics{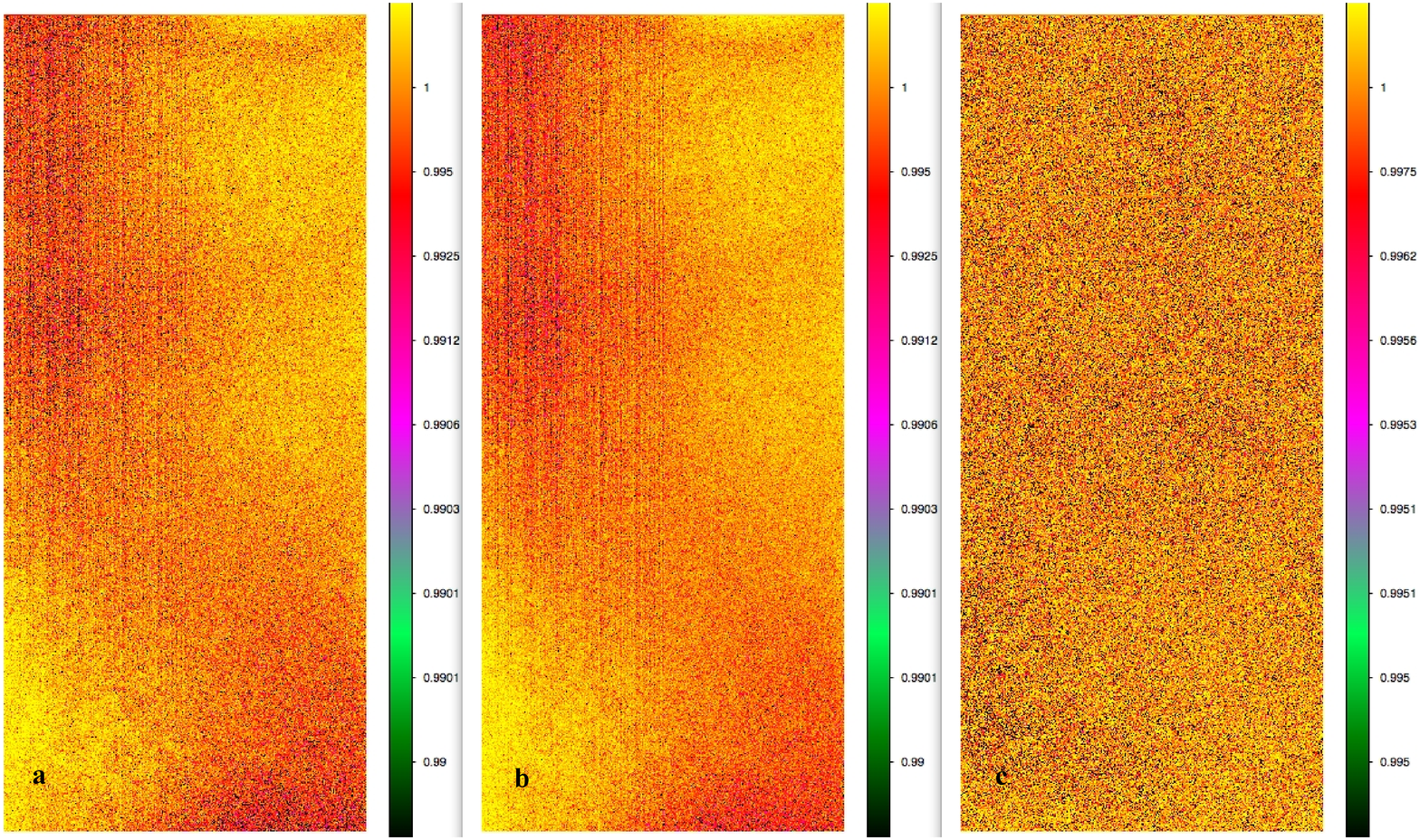}}}
\end{center} {\footnotesize \vspace{1. cm }
Figure 1.  (a), (b) Superflats of the LBC "target" chip obtained in two consecutive nights; (c) superflats ratio showing flats are stable within few days with no specific spatial patterns. 
 \vspace{0.2cm}}

\clearpage

\begin{center}
\vspace{0.cm}
\scalebox{0.4}[0.4]{\rotatebox{0}{\includegraphics{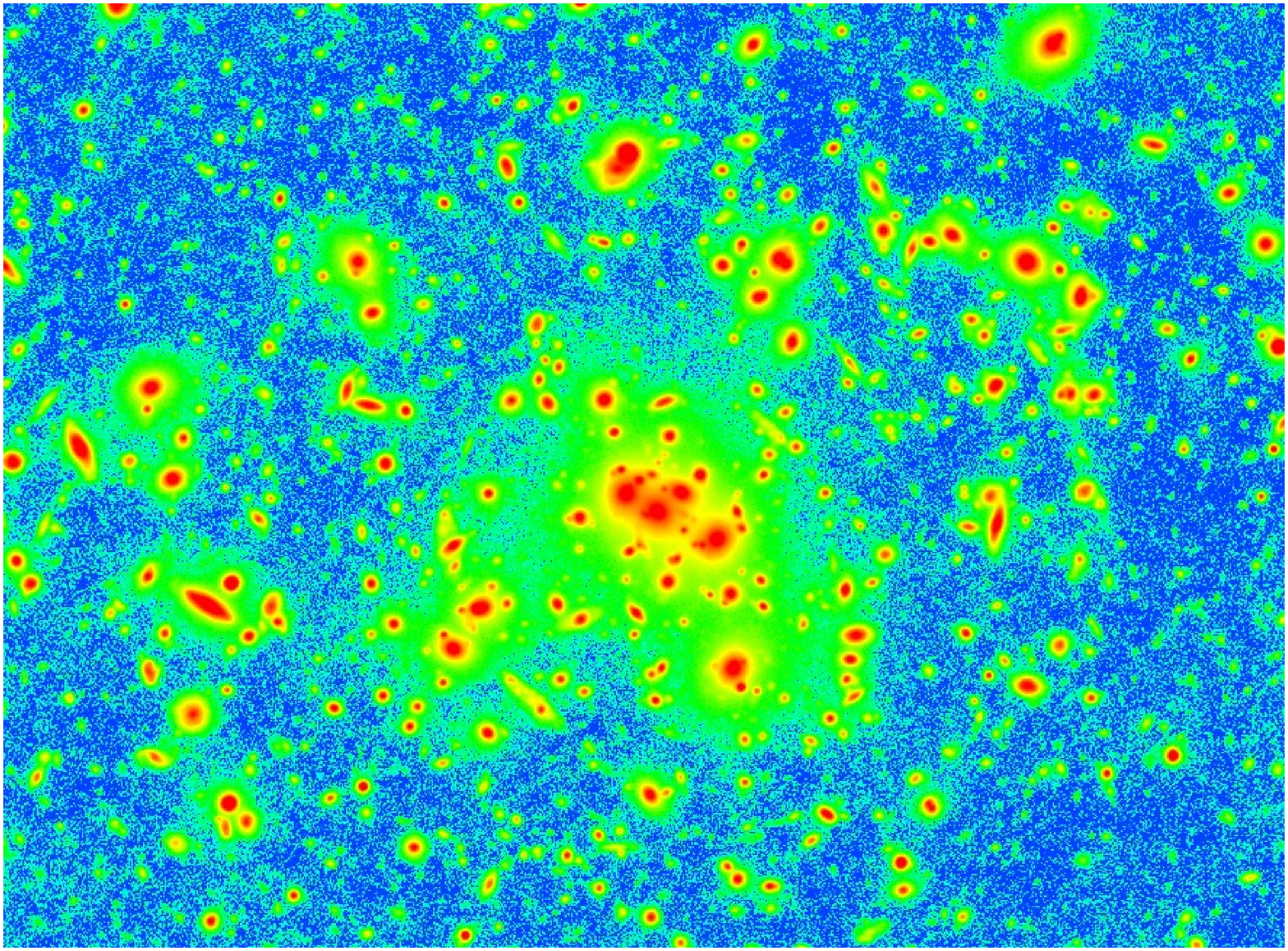}}}
\end{center} {\footnotesize \vspace{1. cm }
Figure 2.  Cluster image in the R band from LBC. The logarithmic intensity scale has been adopted to enlighten the ICL morphology in the inner parts of the cluster. The size of the LBC window shown is about 180$\times$145 arcsec$^2$. The green regions correspond to $R\sim 25.5-26.3$ (at the edge) mag arcsec$^{-2}$. Yellow contours correspond to $R\sim 23.6$ mag arcsec$^{-2}$ and red regions correspond to $R<22.7$ mag arcsec$^{-2}$.
 \vspace{0.2cm}}

\clearpage

\begin{center}
\vspace{0.cm}
\scalebox{0.4}[0.4]{\rotatebox{0}{\includegraphics{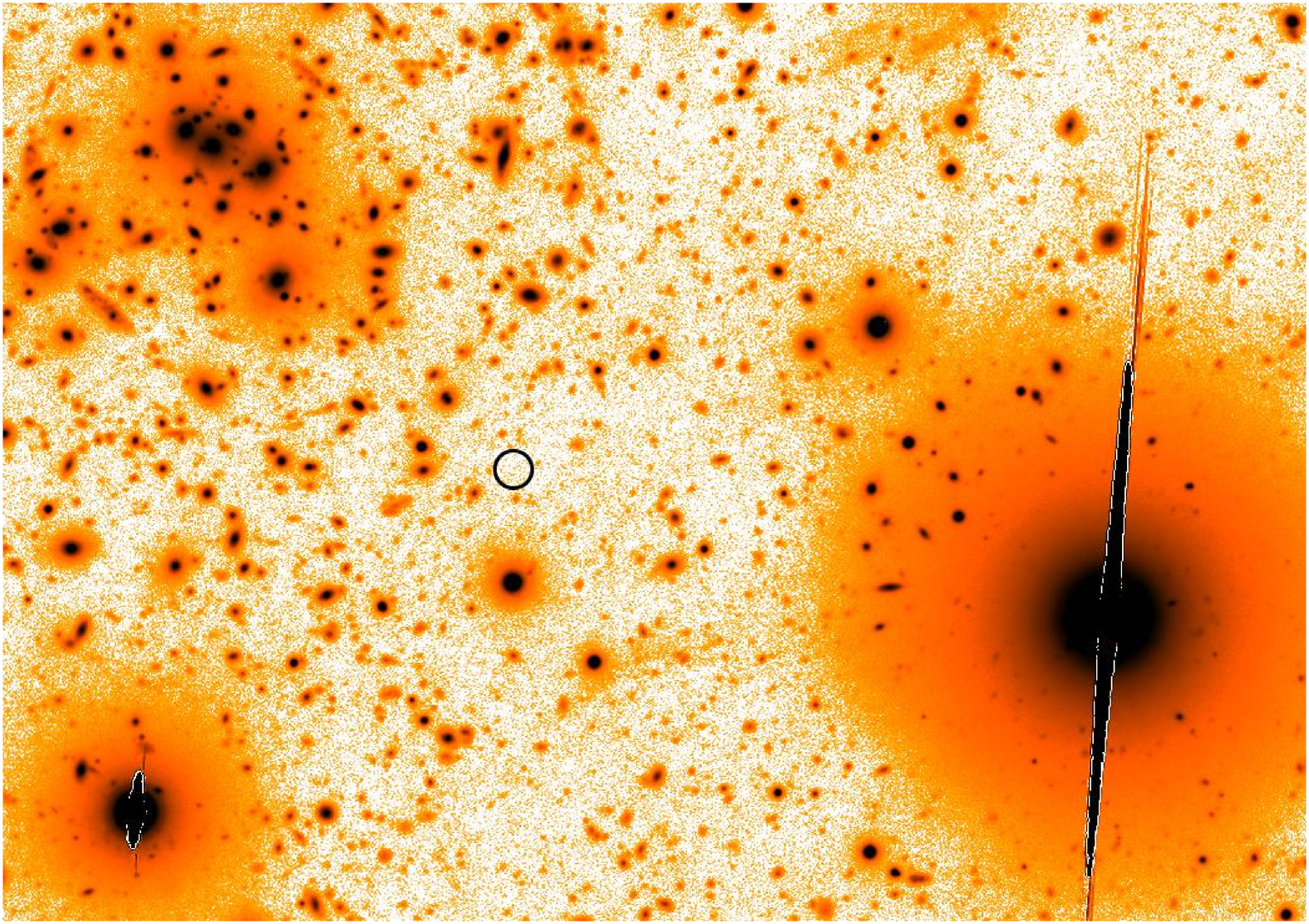}}}
\end{center} {\footnotesize \vspace{1. cm }
Figure 3.  Cluster image in the R band from LBC. The logarithmic intensity scale has been adopted to enlighten the background in the region between the cluster and the nearest brightest stars. The corresponding intensity levels and associated surface brightness are the same as in Figure 2 in the same spatial regions. The small circle denotes the region where the residual background level has been measured as in Figure 4. The size of the image is about 3.9$\times$2.9 arcmin$^2$.
 \vspace{0.2cm}}

\clearpage

\begin{center}
\vspace{0.cm}
\scalebox{0.6}[0.6]{\rotatebox{0}{\includegraphics{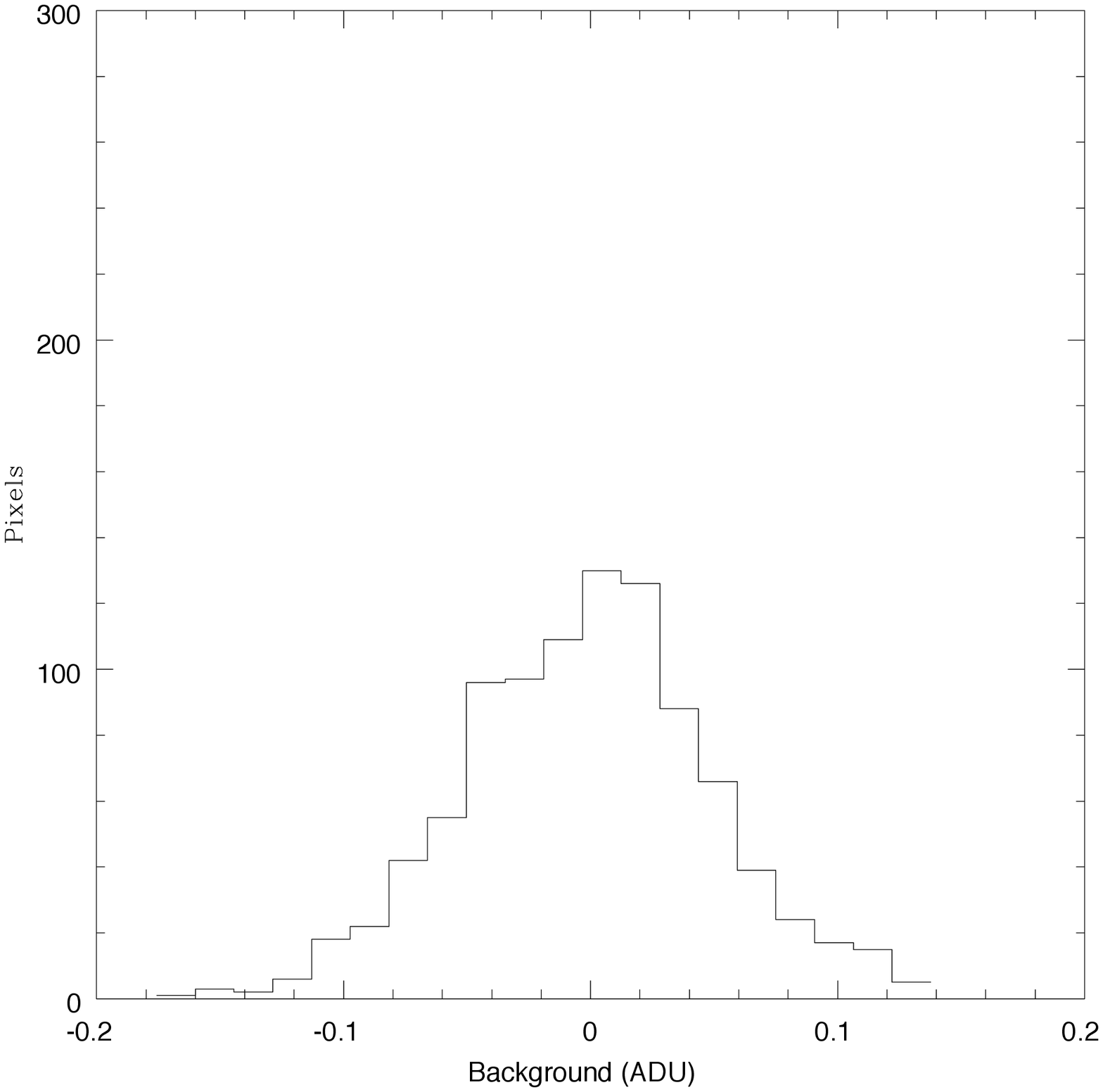}}}
\end{center} {\footnotesize \vspace{0. cm }
Figure 4.  Noise distribution of the background-subtracted image shown in Figure 3. The distribution has been computed inside the small circle. The average value is 0.01 adu with a pixel noise level of 0.06 adu. 
 \vspace{0.2cm}}

\clearpage

\vspace{-0.7cm}
\hspace{-2.5 cm}
\scalebox{0.5}[0.5]{\rotatebox{0}{\hspace{1cm}\includegraphics{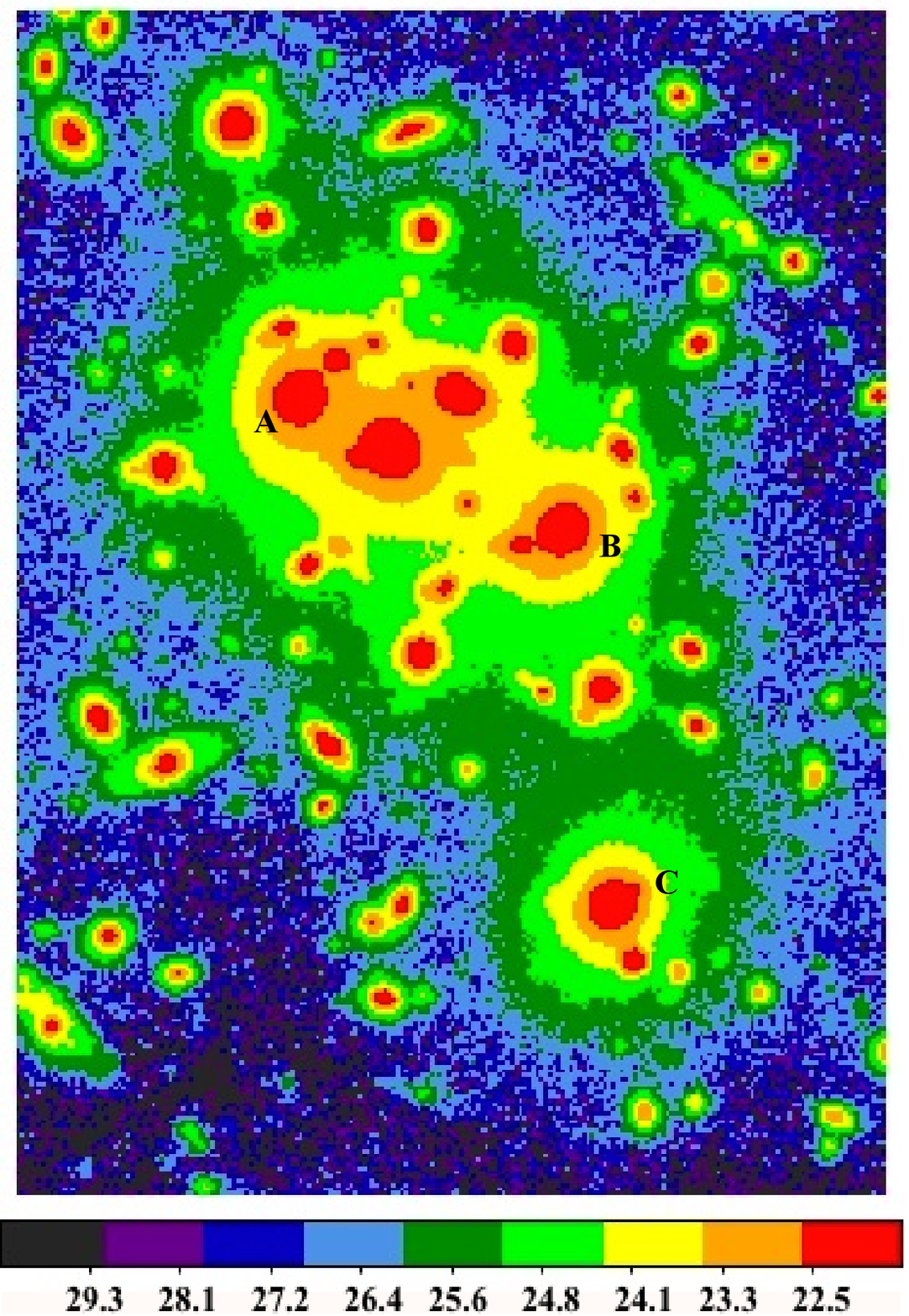}\hspace{1cm}\includegraphics{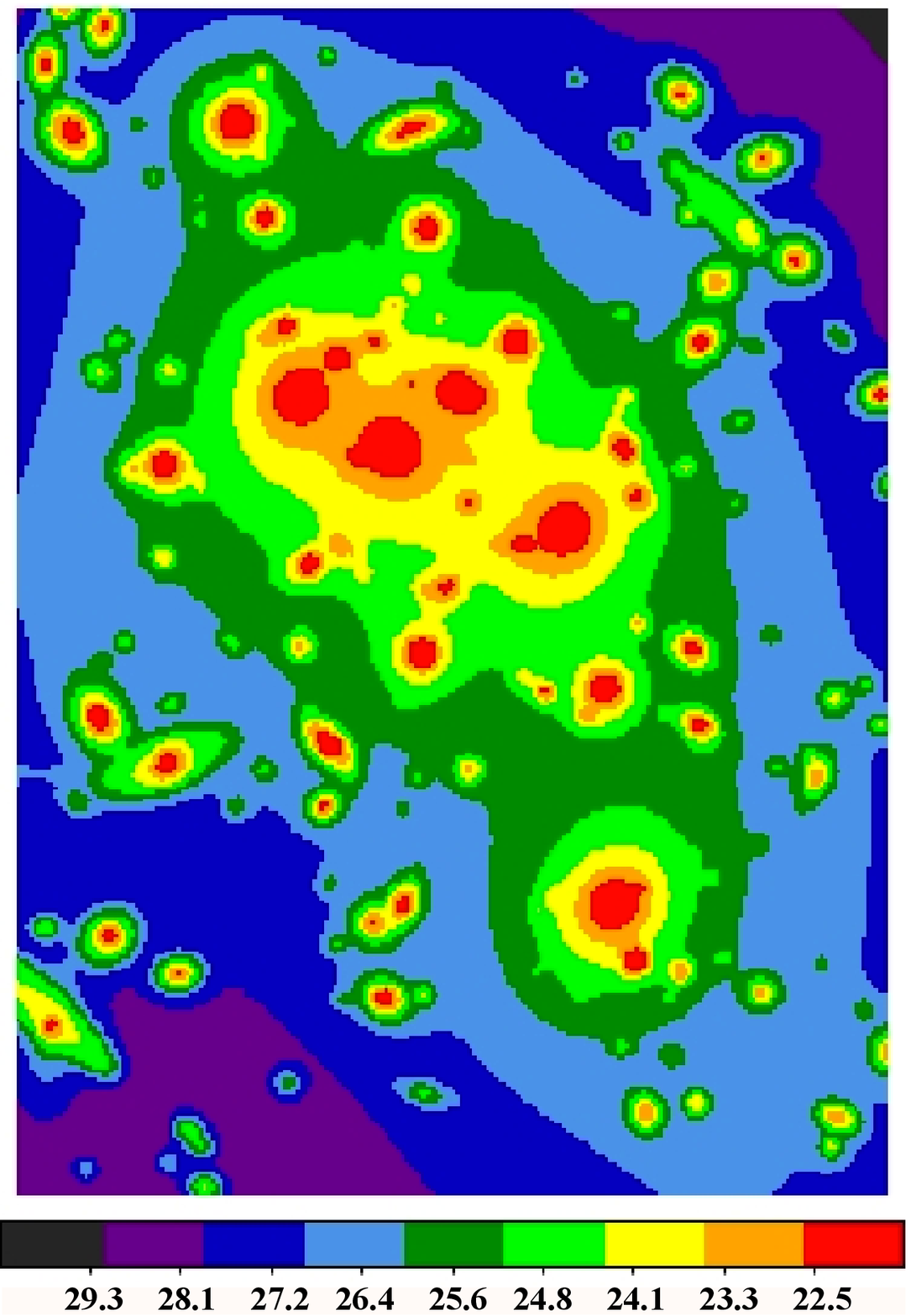}}
\hspace{1cm}\includegraphics{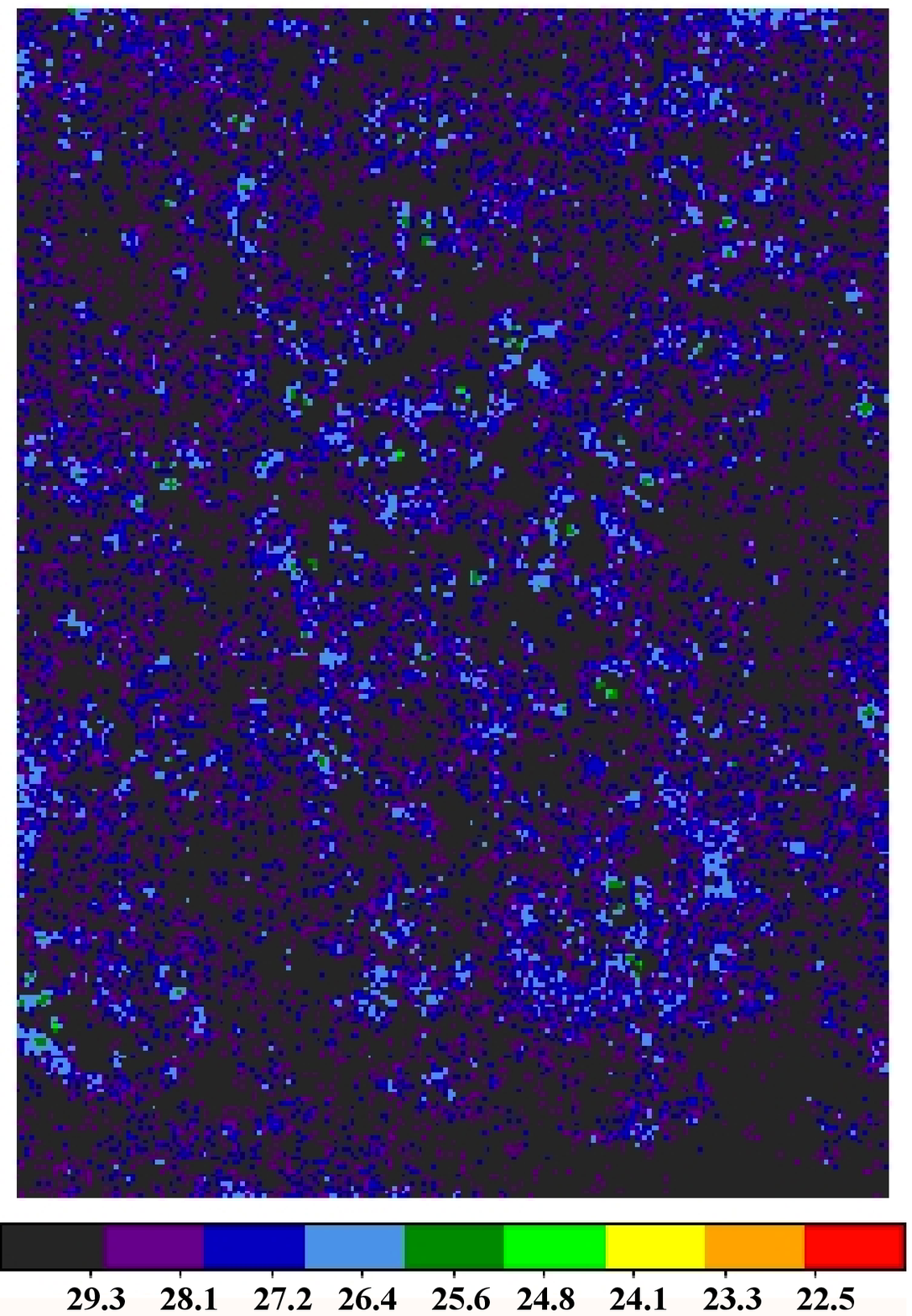}}\\
 {\footnotesize 
Figure 5. (a) Original image of the central fitting region of cluster where all the galaxies have been fitted simultaneously together with a smooth background distribution; galaxies A, B, and C discussed in the text are also shown. The surface brightness values corresponding to the intensity scale are shown at the bottom in mag arcsec$^{-2}$. (b) Image of the best-fit galaxy profiles obtained from galfit, the diffuse background is fitted with a modified Ferrer profile with azimuthal distortion. (c) The residual map derived from galfit; the average value is $-0.001$ with a pixel rms noise of 0.075 adu corresponding to a noise level of about 29 mag arcsec$^{-2}$. The size is  $\sim 46\times 63$  arcsec$^2$.}

\clearpage

\vspace{-0.7cm}
\hspace{-2.5 cm}
\scalebox{0.4}[0.4]{\rotatebox{0}{\hspace{5cm}\includegraphics{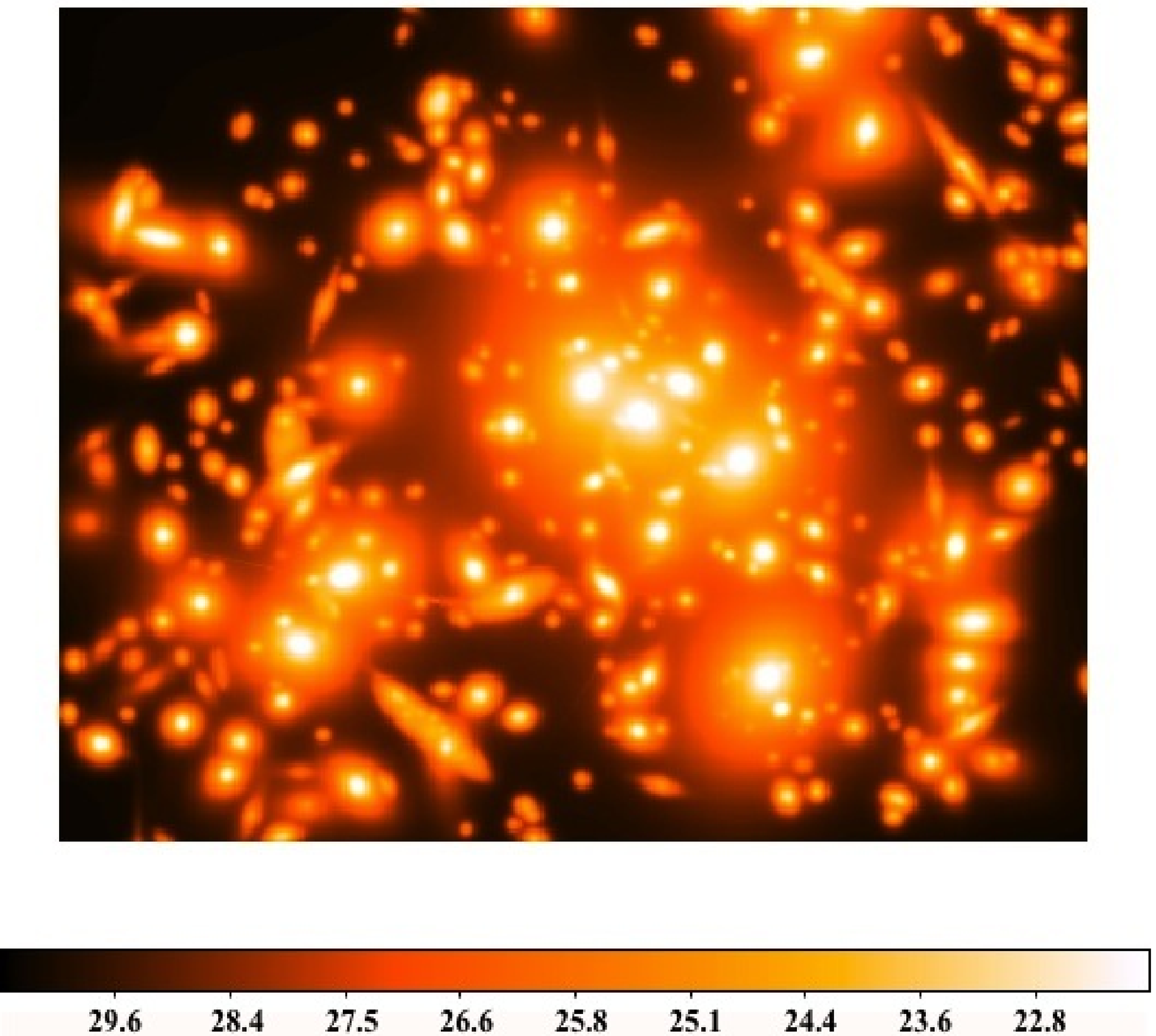}\hspace{5cm}\includegraphics{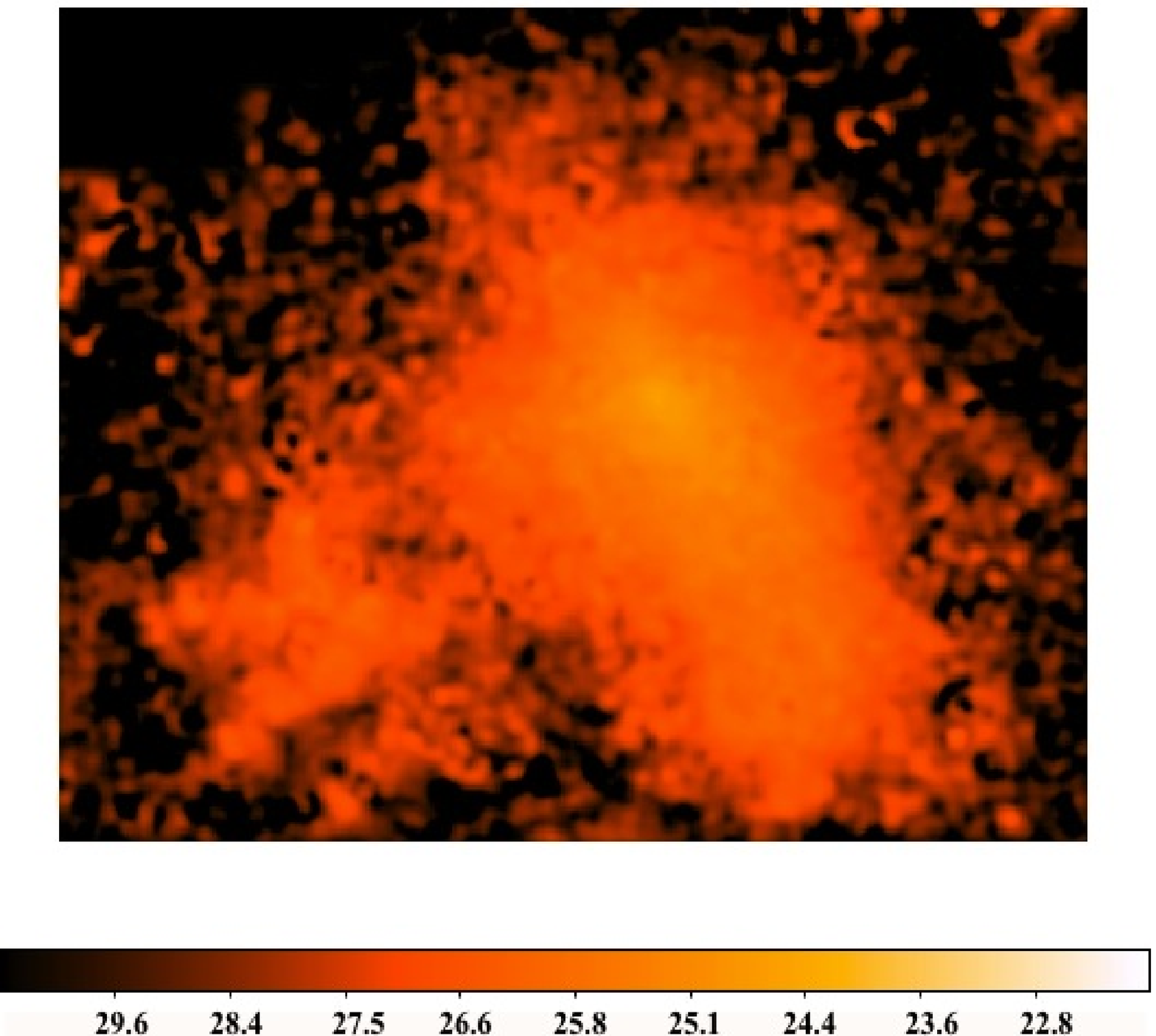}}}
 {\footnotesize 
\\ \\
Figure 6.  (a) Image of the best-fit galaxy profiles obtained from galfit, the diffuse background is not shown and the apparent diffuse light is what is produced only by halos overlap; (b) residual ICL intensity after removal of the galaxy intensities. The surface brightness values corresponding to the intensity scale are shown at the bottom in mag arcsec$^{-2}$. The upper left corner of the image has been left out of the analysis and consequently masked. The size is about 95$\times$77 arcsec$^2$.}

\clearpage

\begin{center}
\vspace{-0.7cm}
\scalebox{0.6}[0.6]{\rotatebox{0}{\includegraphics{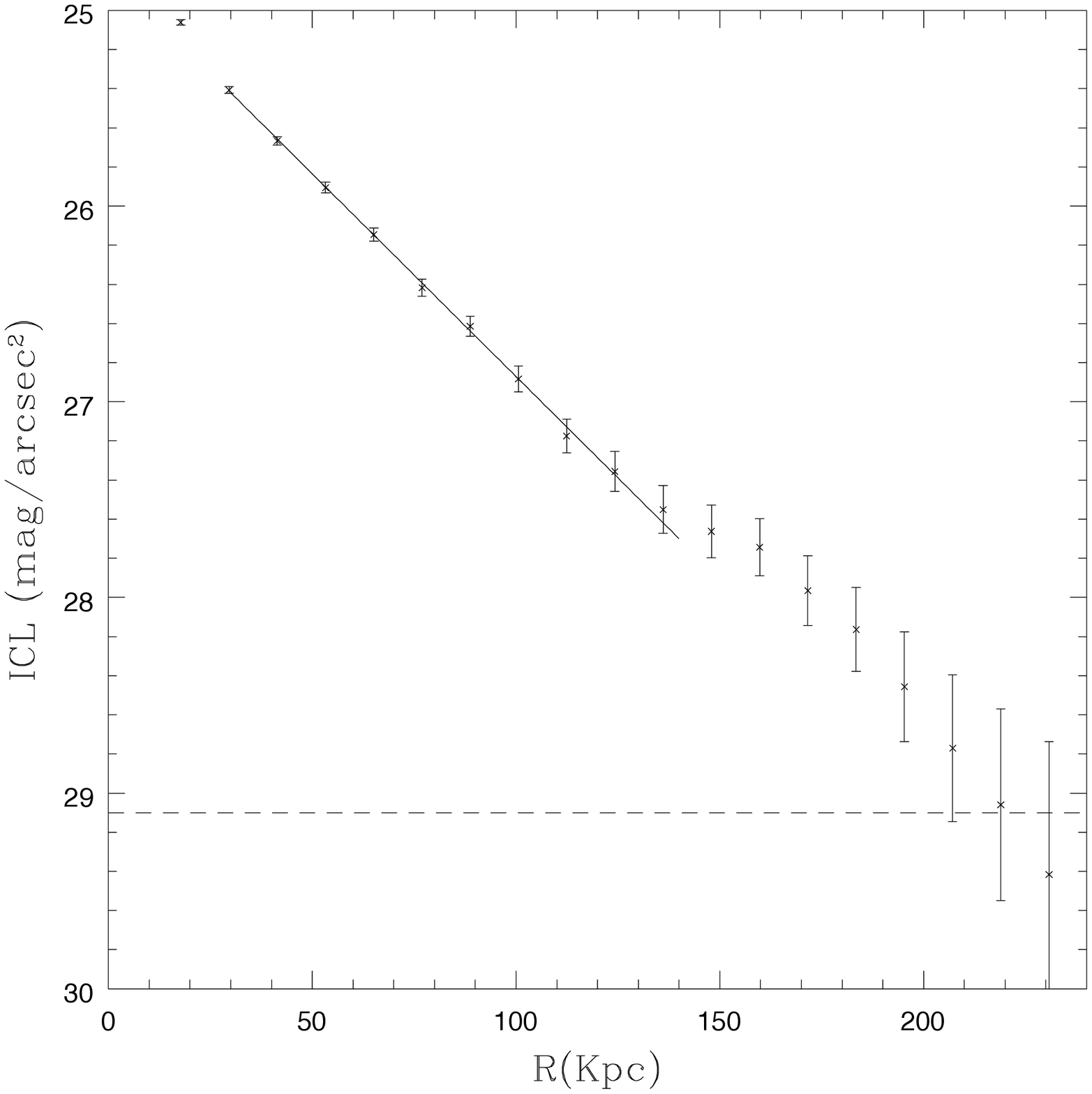}}}
\end{center} {\footnotesize \vspace{-1. cm }
Figure 7.  ICL profile in the red band derived from circular apertures centered on the peak of the ICL surface density. The straight line shows the best fit relation derived in the interval $R=30-140$ kpc. The horizontal line shows the level of the estimated flat fielding noise. Error bars also include the uncertainties in background subtraction (see the text).}
\clearpage

\begin{center}
\vspace{-0.7cm}
\scalebox{0.6}[0.6]{\rotatebox{0}{\includegraphics{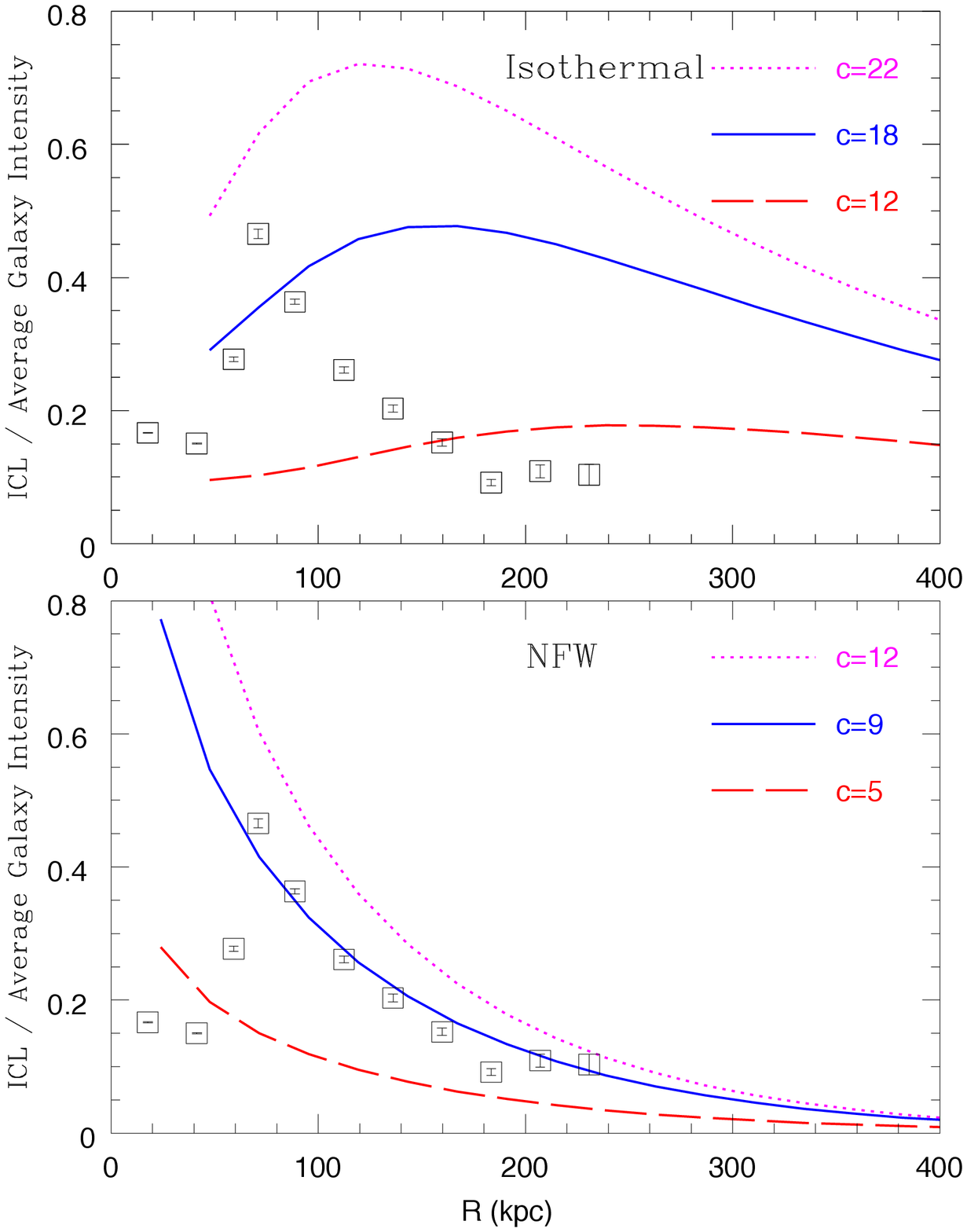}}}
\end{center} {\footnotesize \vspace{1. cm }
Figure 8.  Ratio ${\cal F}_{ICL}$ of the ICL over the integrated galaxy intensity profile computed on circular apertures. Superimposed are theoretical profiles derived from the ratio of the average stellar mass lost from tidal stripping over the total stellar mass assuming in (a) an isothermal profile for the DM potential and in (b) a \citet{Navarro07} profile with different concentration parameters. A good agreement is found adopting a NFW profile with $r_{v}=1.6/h$ Mpc and $c=9$.}
\clearpage

\begin{center}
\vspace{-0.7cm}
\scalebox{0.6}[0.6]{\rotatebox{0}{\includegraphics{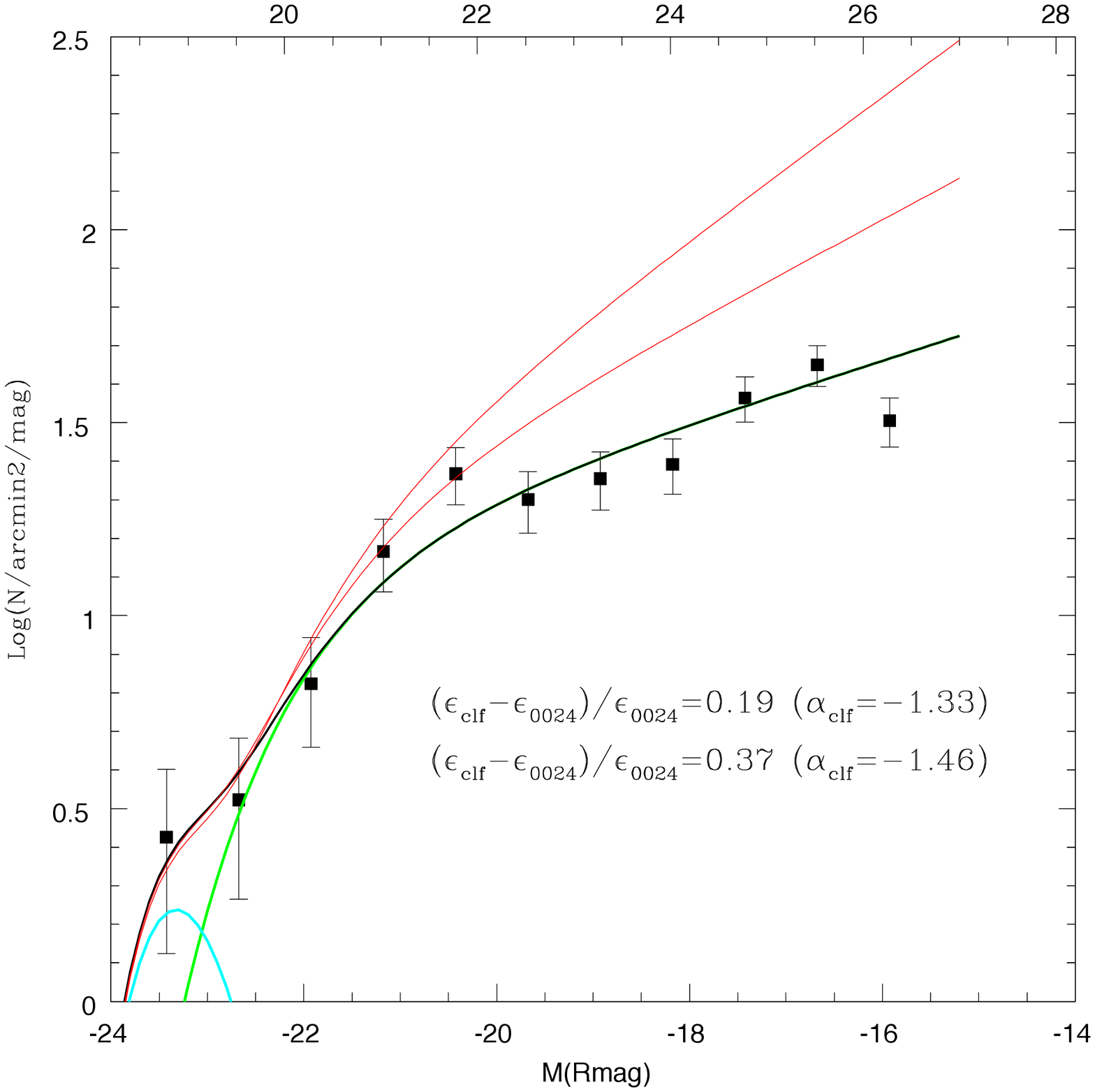}}}
\end{center} {\footnotesize \vspace{-1.5cm }
Figure 9.  Counts in the core of CL0024+17 (black squares and black best-fit, Schechter-type curve) compared with counts derived from cluster luminosity functions obtained from composite samples (red curves). The CL0024 and average cluster curves are derived adopting a double-Schechter shape to take into account the presence of very bright galaxies in the center (cyan curve). The green curve represents the fainter CL0024 Schechter counts. The red curves are derived from composite cluster samples adding the specific bright component of the present cluster (cyan curve). The relative differences in the predicted emissivity $\epsilon$ between the average and the present cluster counts are shown for different faint end slopes bracketing uncertainties in the values derived from the literature. The average counts are normalized to the same $M^*=-22.2$ (see the text for details). The relative differences in the emissivity are of the same order of the average ICL fraction in the same region.}



\begin{thebibliography}{}
\bibitem[Alshino et al. (2010)]{Alshino10}Alshino, A., Khosroshahi H., Ponman, T., Pierre, M., Pacaud, F., \& Smith, G.P. 2010, MNRAS, 401, 941
\bibitem[Arnaboldi (2004)]{Arnaboldi04}Arnaboldi, M. 2004, in IAU Symp. 217, Recycling Intergalactic and Interstellar Matter, ed. P.-A. Duc, J. Braine, \& E. Brinks (San Francisco, CA: ASP), 54
\bibitem[Arnaboldi et al. (2003)]{Arnaboldi03}Arnaboldi, M., Freeman, K. C., Okamura, S. et al. 2003, AJ, 125, 514
\bibitem[Bekki et al. (2003)]{Bekki03} Bekki, K., Couch, W. J., Drinkwater, M. J., \& Shiola, Y. 2003, MNRAS, 344, 399
\bibitem[Bernstein et al. (1995)]{Bernstein95} Bernstein,  G. M., Nichol, R. C., Tyson, J. A., Ulmer, M. P., \& Wittman, D. 1995, AJ, 110, 1507
\bibitem[Bertin \& Arnouts (1996)]{Bertin96} Bertin, E., \& Arnouts, S. 1996, A{\&A}S, 117, 393
\bibitem[Binney \& Tremaine (1987)]{Binney87} Binney, J., \& Tremaine, S. 1987, Galactic Dynamics, (Princeton, NJ: Princeton Univ. press
\bibitem[Boutsia et al. (2011)]{Boutsia11}Boutsia, K., Grazian, A., Giallongo, E., et al. 2011, ApJ, 736, 41
\bibitem[Broadhurst et al. (2000)]{Broadhurst00}Broadhurst, T., Huang, X., Frye, B., \& Ellis, R. 2000, ApJL, 534, L15
\bibitem[Burke et al. (2012)]{Burke12}Burke C., Collins, C. A., Stott, J. P., \& Hilton, M. 2012, MNRAS, 425, 2058
\bibitem[Christlein \& Zabludoff (2003)]{Christlein03}Christlein, D., \& Zabludoff, A. I. 2003, ApJ, 591, 764
\bibitem[Cui et al. (2013)]{Cui14}Cui, W., Murante, G., Monaco, P., Borgani, S. Granato, G. L., Killedar, M., De Lucia, G., Presotto, V., Dolag. K. 2014, MNRAS, 437, 816
\bibitem[Czoske et al. (2001)]{Czoske01}Czoske, O., Kneib, J.-P., Soucail, G., Bridges, T. J., Mellier, Y., \& Cuillandre,
J.-C. 2001, A\&A, 372, 391
\bibitem[Feldmeier et al. (2004)]{Feldmeier04}Feldmeier,  J. J., Ciardullo, R., Jacoby, G. H., \& Durrell, P. R. 2004, ApJ, 615, 196
\bibitem[Fujita (2004)]{Fujita04} Fujita, Y. 2004, PASJ, 56, 29
\bibitem[Fukugita et al. (1995)]{Fukugita95}Fukugita, M., Shimasaku K.,\&  Ichikawa T., 1995, PASP, 107, 945
\bibitem[Gallagher \& Ostriker (1975)]{Gallagher75} Gallagher, J. S., \& Ostriker, J. P. 1972, AJ, 77, 288
\bibitem[Ghigna et al. (1998)]{Ghigna98}Ghigna, S., Moore, B., Governato, F., Lake, G., Quinn, T., \& Stadel, J. 1998, MNRAS, 300, 146
\bibitem[Giallongo et al. (2008)]{Giallongo08} Giallongo, E., Ragazzoni, R., Grazian, A., et al. 2008, A\&A, 482, 349
\bibitem[Gonzalez et al. (2005)]{Gonzalez05} Gonzalez, A. H., Zabludoff, A. I., \& Zaritsky, D. 2005, ApJ, 618, 195
\bibitem[Guennou et al. (2012)]{Guennou12}Guennou, L., Adami, C., Da Rocha, C. et al.  2012, A\&A, 537, 64
\bibitem[Harsono \& DePropris (2009)]{Harsono09}Harsono, D., \& DePropris, R. 2009, AJ, 137, 3091
\bibitem[Henriques \& Thomas (2010)]{Henriques10}Henriques, B. M. B., \& Thomas, P. A. 2010, MNRAS, 403, 768
\bibitem[Jee (2010)]{Jee10}Jee, M. J. 2010, ApJ, 717, 420
\bibitem[Jee et al. (2007)]{Jee07} Jee, M. J., Ford, H. C., Illingworth, G. D. et al. 2007, ApJ, 661, 728
\bibitem[King (1962)]{King62}King, 1962, AJ, 67, 471
\bibitem[Krick \& Bernstein (2007)]{Krick07}Krick,  J. E.,  \& Bernstein, R. A. 2007, AJ, 134, 466
\bibitem[Krick et al. (2006)]{Krick06} Krick, J. E., Bernstein, R. A., \& Pimbblet, K. A. 2006, AJ, 131, 168
\bibitem[Martel et al. (2012)]{Martel12}Martel, H., Barai, P., \& Brito, W. 2012, ApJ, 757, 48
\bibitem[Matthews et al. (1964)]{Matthews64} Matthews, T. A., Morgan, W. W., \& Schmidt, M. 1964, ApJ, 140, 35
\bibitem[Merritt (1983)]{Merritt83} Merritt, D. 1983, ApJ, 264, 24
\bibitem[Mihos (2004)]{Mihos04} Mihos, J. C. 2004, in IAU Symp. 217, Recycling Intergalactic and Interstellar Matter, ed. P.-A. Duc, J. Braine \& E. Brinks (San Francisco, CA: ASP),  390
\bibitem[Mihos et al. (2005)]{Mihos05}Mihos, J. C., Harding, P., Feldmeier, N., \& Morrison, H. 2005, ApJL, 632, L41
\bibitem[Miller \& Smith (1983)]{Miller83} Miller, R. H., Smith, B. F. 1983, Internal Kinematics and Dynamics of Galaxies,  E. Athanassoula Ed., Proc. SPIE, 100, 351
\bibitem[Mo, Mao, White (1998)]{Mo98}Mo, H. J., Mao, S., \& White, S. D. M. 1998, MNRAS, 295, 319
\bibitem[Moore et al. (1996)]{Moore96} Moore, B., Katz, N., Lake, G., Dressler, A., \& Oemler, A. 1996, Natur, 379, 613
\bibitem[Moran et al. (2005)]{Moran05} Moran,  S. M., Ellis, R. S., Treu, T., Smail, I. R., Dressler, A., Coil, A., \& Smith, G. P. 2005, ApJ, 633, 32
\bibitem[Navarro, Frenk, \& White (1997)]{Navarro97}Navarro, J. F., Frenk, C. S., \& White, S. D. M. 1997, ApJ, 490, 493
\bibitem[Peng et al. (2010)]{Peng10}Peng, C. Y., Ho, L. C., Impey, C. D., \& Rix, H.-W. 2010, AJ, 139, 2097
\bibitem[Puchwein et al. (2010)]{Puchwein10}Puchwein, E., Springel, V., Sijacki, D., \& Dolag, K. 2010, MNRAS, 406, 936
\bibitem[Richstone (1975)]{Richstone75} Richstone, D. O. 1975, ApJ, 200, 535
\bibitem[Rudick et al. (2011)]{Rudick11}Rudick, C. S., Mihos, J. C., \& McBride, C. K. 2011, ApJ, 732, 48
\bibitem[Seigar et al. (2007)]{Seigar07} Seigar, M. S., Graham, A. W., \& Jerjen, H. 2007, MNRAS, 378, 1575
\bibitem[Shombert (1988)]{Shombert88} Shombert, J.M., 1988, ApJ, 328, 475
\bibitem[Smith et al. (2005)]{Smith05} Smith, G. P., Treu, T., Ellis, R. S., Moran, S. M., \& Dressler, A. 2005, ApJ, 620, 78
\bibitem[Taylor \& Babul (2001)]{Taylor01}Taylor, J. E., \& Babul, A. 2001, ApJ, 559, 716
\bibitem[Toledo et al. (2011)]{Toledo11}Toledo, I., Melnick J., Selman F., Quintana H., Giraud E.,
\& Zelaya P., 2011, MNRAS, 414, 1
\bibitem[Treu et al. (2003)]{Treu03}Treu, T., Ellis, R. S., Kneib, J., Dressler, A., Smail, I., Czoske, O., Oemler, A., \& Natarajan, P. 2003, ApJ, 591, 53 
\bibitem[Tutukov \& Fedorova (2011)]{Tutukov11} Tutukov, A. V., \& Fedorova, A. V.  2011, Astron. Rep., 55, 383
\bibitem[Tyson et al. (1998)]{Tyson98} Tyson, J. A., Kochanski, G. P., \& dell’Antonio, I. P. 1998, ApJL, 498, L107
\bibitem[Umetsu et al. (2010)]{Umetsu10} Umetsu, K., Medezinski, E., Broadhurst, T., et al. 2010, ApJ, 714, 1470
\bibitem[von Hoerner (1957)]{vonh57} von Hoerner, S. 1957, ApJ, 125, 451
\bibitem[Weil et al. (1997)]{Weil97}Weil M. L., Bland-Hawthorn, J., \& Malin, D. F. 1997, ApJ, 490, 664
\bibitem[Zibetti et al. (2005)]{Zibetti05}Zibetti S., White, S. D. M., Schneider, D. P., \& Brinkmann, J. 2005, MNRAS, 358, 949
\bibitem[Zwicky (1951)]{Zwicky51}Zwicky F. 1951, PASP , 63, 61
\end{thebibliography}
\end{document}